\documentclass[a4paper,12pt]{article}
\usepackage[utf8]{inputenc}
\usepackage{amsmath,amssymb}
\usepackage{graphicx}
\usepackage{float}
\usepackage{xcolor}
\usepackage[colorlinks=true, linkcolor=magenta, citecolor=red, urlcolor=black]{hyperref}
\usepackage{fancyhdr}
\allowdisplaybreaks
\usepackage{orcidlink}  

\begin{document}

\begin{center}
    {\LARGE \bfseries Bouncing Cosmologies in Modified Gravity with Spacetime Torsion \par}
    \vskip 1em
    {\large
    Sonej Alam$^{1}$\thanks{e-mail: sonejalam36@gmail.com}{\orcidlink{0009-0008-8322-2923}},
    Somasri Sen$^{1}$\thanks{e-mail: ssen@jmi.ac.in},
    Soumitra Sengupta$^{2}$\thanks{e-mail: tpssg@iacs.res.in}
    \par}
    \vskip 1em
    {\small
    $^{1}$Department of Physics, Jamia Millia Islamia, New Delhi 110025, India \\
    $^{2}$School of Physical Sciences, Indian Association for the Cultivation of Science, Kolkata 700032, India
    }
\end{center}
\vskip 1em

\thispagestyle{fancy}
\fancyhf{} 
\fancyfoot[C]{\footnotesize\textit{E-mails:} 
\href{mailto:sonejalam36@gmail.com}{sonejalam36@gmail.com}, 
\href{mailto:ssen@jmi.ac.in}{ssen@jmi.ac.in}, 
\href{mailto:tpssg@iacs.res.in}{tpssg@iacs.res.in}}
\renewcommand{\headrulewidth}{0pt}
\renewcommand{\footrulewidth}{0pt}

\noindent

\begin{abstract}
We explore the possibility of realizing a non-singular bounce in the early universe within the framework of modified gravity with spacetime torsion. In Einstein Cartan theory, torsion is embedded in the spacetime by adding an antisymmetric part in affine connection . We consider generalized version of the framework as $f(\bar{R})$, $\bar{R}$ being the scalar of the modified curvature tensor. $f(\bar{R})$ gravity is recast in Einstein frame as non-minimally coupled scalar tensor theory  where the scalar field gets coupled with a rank 2 antisymmetric torsion field through derivative couplings. We investigate whether the introduction of three additional torsion-dependent terms in  Einstein frame help to realize a bounce. We first explore this cosmological system in the background of a homogeneous and isotropic FRW spacetime but inclusion of the torsion terms are insufficient to produce a bounce in this symmetric setting. Motivated by this limitation, we relax the symmetry and generalize the background to include inhomogeneity and anisotropy. In this  setup, the dynamics is modified in such a way that a bouncing solution is possible without invoking phantom fields or energy condition violations. We have found the exact solutions of all the fields and reconstructed the modified gravity form. We have addressed the behaviour of the fields under perturbation and investigated the stability of the solutions. Constraints on the model parameters have also been derived based on cosmological observations.
\end{abstract}


\section{Introduction}
The diverse cosmological observations of recent decades (Planck\cite{Planck2018}, JWST\cite{2006SSR},  DESI\cite{225ar}, Euclid\cite{2025arXiv250315332E}, DES\cite{2024ApJ...973L..14D}, ACT\cite{2025arXiv250314454C} and others) have confirmed the eminence of cosmology in the contemporary period.  The facts like we have a flat universe which has a faster expansion rate at present and 13.7 billion years ago, an intermediate stage of slower expansion forming large scale structures and contains mostly dark matter and dark energy are all well established and substantiated by observations\cite{1998AJ....116.1009R,PhysRevLett.83.670,2004PhRvD..69j3501T,2003ApJS..148..175S,Abazajian_2004,2013ApJS..208...20B,PhysRevD.86.103518}. However, in spite of these achievements, many essential questions continue to be unanswered; specifically, the nature of dark energy and dark matter is still mysterious, the cause of acceleration at two extreme epochs (early and late) is not known, and we also lack definitive insights regarding the universe's initial conditions and the final outcome\cite{Bamba:2022jyz,2022JHEAp..34...49A}. 

The standard Big Bang model based on the foundation of General Relativity (GR) has effectively addressed many intriguing issues of the universe ranging from the perihelion shift of Mercury, the deflection of light by the Sun,  expansion of the universe to gravitational lensing, gravitational waves, black holes, etc. However, despite its success, it encounters hurdles both in early and late universes. In late universe, Cosmological Constant $\Lambda$, is needed in the model, which is just added to the theory to justify observations. But, inclusion of $\Lambda$ not only has theoretical concerns\cite{2006IJMPD..15.1753C}, but also observational challenges\cite{2022ApJ...937L..31S,2022JHEAp..34...49A,2022NewAR..9501659P,2021APh...13102605D,2025arXiv250401669D,2024PhRvD.110l3518P,2023ARNPS..73..153K,2021APh...13102604D,2022JHEAp..34...49A,2021MNRAS.505.5427N,DiValentino:2018gcu}. The early universe issues include initial singularity, flatness problem, horizon problems among others. Though inflation is successful in answering  horizon problem and flatness problem and even generating nearly scale-invariant primordial fluctuations consistent with the Planck observations\cite{BROUT197878,PhysRevD.23.347,STAROBINSKY198099}, however if we extrapolate further into past we hit a curvature singularity, where all the fundamental laws of physics, break-down\cite{Friedman:1922kd,Friedmann:1924bb}. These issues of early and late universe prompted modification of GR in various ways. A prominent one is $f(R)$ theory with different functions of $R$ in contrast to only $R$ in Einstein theory\cite{2004PhRvD..70d3528C,2003PhRvD..68l3512N,2007IJGMM..04..115N,2011PhR...505...59N,doi,DeFelice:2010aj,2010RvMP...82..451S,2024arXiv240302771A}.  Higher powers of $R$ in these models can explain early-stage accelerated expansion known as inflation, while exploring negative powers of $R$ offer insights into late-time acceleration in the universe\cite{2011PhR...509..167C,2012Ap&SS.342..155B,PhysRevD.75.083504,2022EPJC...82..264B}. Other modifications of Einstein gravity include $f(\cal{T})$\cite{PhysRevD.81.127301},$f(Q)$\cite{2017arXiv171003116B}, $f(G)$\cite{2007JPhCS..66a2005N}, $f (R, T)$\cite{PhysRevD.84.024020}, $f(R,G)$\cite{2005PhLB..631....1N}, $f(Q,T)$\cite{2019EPJC...79..708X}, $f (R, L_m)$\cite{2010EPJC...70..373H}, ($\cal{T}$:Torsion scalar, $Q$:nonmetricity scalar, G: Gauss-Bonneett invariant, T: trace of energy momentum tensor, $L_m$: matter lagrangian) among others.

The singularity at the initial time remains one of the most fundamental unresolved problems in theoretical cosmology and till date it has not been resolved\cite{Friedman:1922kd,Friedmann:1924bb}. This may be due to the attractor nature of inflation where the information on the initial singularity was lost as the exponential expansion stretched away the initial spatial curvature. To overcome the singularity problem, various alternative cosmological models have been proposed\cite{2008PhR...465..223L,2008PhR...463..127N,2014SCPMA..57.1414C}. One approach is bouncing cosmology, where the universe contracts to a minimum size and then transitions into an expansion phase, thus replacing the initial singularity with a smooth bounce\cite{2015PhR...571....1B}.
Under the purview of GR, the bouncing behavior cannot be achieved since it leads to the violation of the null energy condition, which is happening because of the change in sign of the expansion rate. To achieve this, new physics is required to supply the bounce, which can be obtained either by introducing new kinds of matter such
as phantom or quintom fields \cite{2007JHEP...10..071C}, ghost condensates\cite{PhysRevD.66.063509,2011JCAP...04..019L}, effective string theory actions\cite{PhysRevD.67.124003}, Galileons\cite{2011JCAP...10..036Q}, nonminimal coulings\cite{2025PhyU...68..163V,2020PhRvD.101d3506N} or by
using modified theories of gravity such as $f(R)$\cite{2014JCAP...01..008B,Barragan:2009sq,2010PhRvD..82h4015B}, $f(G)$\cite{2014PhLB..732..349B,Bamba_2015}, $f(\cal{T})$\cite{2011CQGra..28u5011C}, $f(Q)$\cite{2023EPJC...83..113A}, even sometimes coupled with other like $f(R, T)$\cite{PhysRevD.97.123536,2019EPJP..134..504T,Mishra:2019oul}, $f(Q, T)$gravity \cite{AGRAWAL2021100863}, $f(R, G)$\cite{2022Univ....8..636L}, $f(R,G,T)$\cite{2025arXiv250320709M}.
Different approaches have been adopted to avoid the initial singularity. One popular approach is to take different functional forms of the scalars like Ricci scalar $R$\cite{2024PhyS...99f5031J,2023GrCo...29..294A}, Torsion scalar$\cal{T}$\cite{2024EPJC...84.1202T,2024arXiv241004062D,2024arXiv240615825W}, non-metricity scalar $Q$\cite{2024EPJP..139..799K,2025arXiv250421757M}, Gauss-Bonnet invariant $\cal{G}$\cite{2022Univ....8..636L} and others. Most of these investigations are carried out in Jordon frame where gravity is modified\cite{2024arXiv240615825W,2025arXiv250421757M,2025arXiv250320709M}.For comprehansive review of bouncing cosmology formulated within the framework of modified theories of gravity and other theories, the reader may refer to \cite{Nojiri:2017ncd,2017FoPh...47..797B,2015PhRvD..92b4016O,2015PhR...571....1B,2008PhR...463..127N} .

Among the modified gravitational theories, the generalized version of Eintein Hilbert action with higher order curvature term, popularly known as $f(R)$ theories, has garnered significant attention. The inclusion of higher curvature terms in the action typically results in the appearance of ghosts from higher derivative terms, which in turn leads to instability. A conformal transformation of metric can convert $f(R)$ theories into a scalar-tensor theory in Einstein frame at the action level\cite{2011PhR...505...59N,2011PhR...509..167C}. Then, the instability issue in the original theory reflects in the  the kinetic and potential terms of the scalar field, where the scalar field  has a stable minima in the potential and a kinetic term with proper signature.

Torsion appears naturally in theories like Einstein–Cartan gravity
and allows the spacetime connection to include antisymmetric parts related to spin and intrinsic angular momentum\cite{RevModPhys.48.393,2002PhR...357..113S}. When the connection is modified to combine torsion with gravity, the dynamics changes, offering new degrees of freedom that opens up new possibilities. In this context a spin one, rank 2 antisymmetric tensor field, better known as massless Kalb-Ramond field has been extensively investigated for different aspects\cite{2001PhLB..508..381D,PhysRevD.98.104002,PhysRevD.9.2273,Mukhopadhyaya:2001dp,2002EPJC...23..357K,2024EL....14729001K,Kar:2001eb,2002PhRvL..89l1101M,2016AnPhy.367..258C}. 

It has been shown in \cite{2024EL....14729001K} that a generalized form of Einstein Cartan theory with antisymmetric part in affine connection can be mapped into a non-minimally coupled scalar-tensor theory with a scalar and a rank-2 massless antisymmetric tensor field which can be further expressed in the Einstein frame, using conformal transformation of metric, similar to $f(R)$ theories without torsion\cite{2011PhR...505...59N,2011PhR...509..167C}. The issue of instability is also addressed in \cite{2024EL....14729001K}
Thus, the action of a generalized Einstein Cartan theory ($f({\bar{R}})$) is expressed in Einstein frame as a scalar–tensor theory,  where the scalar field gets non-minimally coupled with the rank-2 massless antisymmetric field through derivative couplings\cite{2024EL....14729001K}. Conformal transformations are reparameterization of a metric which changes the background equations and hence the dynamics, though the perturbation remains invariant for a single scalar field\cite{PhysRevD.99.103532}.  Einstein frame are commonly referred as physical frame  as observations are made in Einstein frame. 

The present work explores the feasibility of obtaining an exact bouncing solution in a generalized form of Einstein Cartan theory $f({\bar{R}})$ recast in Einstein frame, through conformal transformation. The present investigation is carried out in both isotropic and anisotropic spacetime.
However, even with the non minimal derivative coupling  between scalar and torsion  tensor, we found that within the FRW framework it is difficult to realize a bounce without violating null energy conditions. This investigation reveals that assuming homogeneity and isotropy may be too restrictive, pointing to the limitations of such symmetry. Motivated by this, we explore the possibility of achieving a bounce by relaxing the symmetry assumptions of the metric. Research on bouncing cosmology in anisotropic spacetimes, such as Bianchi cosmology, can be found in the existing literature \cite{2022PhyS...97b5002S,2022ForPh..70a0065A,2018PhRvD..98b4009C}. 

In Section 2, we review the conditions of bouncing cosmology in the General  Relativistic framework and detail the framework of the generalized form of Einstein Cartan gravity on which the study is based.  We mention briefly how the $f(\bar{R})$ gravity is formulated and then give a brief discussion on the conformal mapping which expresses the theory as a non-minimally coupled scalar-tensor action in Einstein frame. In section 3, we workout the dynamics of evolution in the background of a isotropic and homogeneous spatially flat Friedmann–Robertson–Walker (FRW) universe and analyze the necessary conditions for a bouncing solution within this setup. We proceed with our study by including three different terms in the action systematically one by one according to their coupling strength in three different cases. We derive the modified field equations in the presence of torsion and re-examine the conditions for achieving a bounce. Despite the additional degrees of freedom introduced by torsion, we find that a non-singular bounce still cannot be realized in the FRW metric without violating null energy conditions. This is a general conclusion without assuming any particular functional form of $f({\bar{R}})$. Our investigation indicates that the symmetry of the spacetime might be too restrictive to realize bounce. Motivated by this limitation, in Section 4, we relax the symmetry in the spacetime and consider a slightly different cosmological background to explore whether a bounce solution can emerge naturally in such background, without violating the standard null energy constraints. Within this framework, it is possible to realize a non-singular bouncing solution while maintaining consistency with the null energy condition. This is also achieved without any functional form $f(\bar R)$. We find the exact solutions for the fields and reconstruct the form  of higher curvature.We also examine briefly the stability of the solutions and constrained the model parameters using cosmological observations. A conclusion follows section 5.

 \section{ Background and formulation}
The standard Big Bang model has two major issues in the early universe : the epoch right after the bang, which is confronted with Inflationary cosmology and the bang itself which is addressed in Bouncing cosmology. This work deal with bouncing scenario. So, lets summarize the conditions based on which the bouncing scenarios are developed.
\begin{itemize}
   
\item In a bouncing scenario, the scale factor $a$ behaves such that it contracts or decreases with time till it reaches a non-zero critical value at a time identified as bouncing epoch i.e before the bouncing epoch $\dot{a}<0$ or $H\equiv\frac{\dot{a}}{a}<0$
\item
After the bounce, the scale factor  expands or increases with time i.e after the bouncing epoch $\dot{a}>0$ or $H\equiv\frac{\dot{a}}{a}>0$ expansion.
 
 \item At the bouncing epoch, the scale factor $a$ posses a non-zero finite critical value, resulting in vanishing Hubble parameter $H=\frac{\dot{a}}{a}=0$ and diverging deceleration parameter $q=-\frac{1}{H^2}\frac{\ddot{a}}{a}$. 

 \item 
 At the bouncing epoch the scale factor attains a minimum critical value or $H$ vanishes which implies $\dot{H}$ must be positive at the time of bounce.
 
 \item 
The Ray Chaudhury equation, in the context of General Relativity (GR) predicts the Null Energy condition (NEC) to be $\rho+p\geq 0$ in terms of perfect fluid distribution of matter. For a FRW universe, $\dot{H}=-4\pi G(\rho+p)$, which implies that in the vicinity of the bouncing epoch, the NEC is violated.
\item 
This condition rules out the scenario in the context of General Relativity (GR) with FRW background, unless the model has phantom or ghost behaviour.
\item 
The $\dot{H}$ equation also indicates that, to realize a bouncing model, the equation of state parameter $\omega$ crosses the phantom divide twice in the vicinity of the bounce epoch (before and after the bounce).
\end{itemize}

These points sum up the fact that it is not possible to realize a bounce in the context of General Relativity with a FRW background with canonical fields or matter distribution. Hence different kind of approaches has been adopted out of this framework, which is detailed in the introduction. In this article we investigate bouncing solution in generalized Einstein Cartan theory $f(\bar R)$ without assuming a functional form of the scalar $\bar{R}$ in the theory. Further the investigation is carried in Einstein frame, which is often treated as physical frame.  First, we summarize how the generalized Einstein Cartan theory $f(\bar R)$ is formulated and then map it in the Einstein frame.


We discuss briefly the formulation of a generalized Einstein-Cartan action $f(\bar{R})$, the details of which can be found in \cite{2024EL....14729001K}. Torsion is added to the curvature by adding the antisymmetric part of the affine connection in defining the Riemann tensor. Then the higher curvature is constructed with the modified Riemann tensor. 

A more general Christoffel connection can be defined by metric tensor and contorsion tensor using metric compatibility as
$$\bar{\Gamma^d}_{ab} = {{\Gamma}^d}_{ab} - {K^d}_{ab},$$
where ${{\Gamma}^d}_{ab}$ is symmetric in the lower indices 
$${{\Gamma}^d}_{ab}=\frac{1}{2}g^{dc}\left(\partial_ag_{eb}+\partial_bg_{ea}-\partial_eg_{ab} \right)$$
and the contortion tensor ${K^d}_{ab}$ defined in terms of torsion tensor \({\cal{T}}^{a}_{bc}\) as
$$
    {K^d}_{ab} = \frac{1}{2} \left( {{{\cal{T}}^d}_{ab}} - {{{{\cal{T}}_a}^d}_b} - {{{{\cal{T}}_b}^d}_a} \right)
$$
${K^d}_{ab}$ is antisymmetric in the first two indices, i.e., \(K_{abc} = -K_{bac}\), while the torsion tensor \({\cal{T}}^{a}_{bc}\) is antisymmetric in the last two indices.


In terms of the new Christoffel connection, the Riemann tensor can be expressed as function of symmetric Christoffel connections and the contorsion tensor as:
\begin{equation}
     \bar{R}_{bd}=R_{bd}-{\nabla}_a {K^a}_{bd}+{\nabla}_d {K^a}_{ba}+{K^a}_{ea}{K^e}_{bd}-{K^a}_{ed}{K^e}_{ba}~~.
     \label{Ricci}
\end{equation}
The modified Ricci tensor $\bar{R}_{bd}$ is sum of the Ricci tensor built from the symmetric part of the Christoffel connections and derivative of contorsion and higher order terms in it. The covariant derivatives in (\ref{Ricci}) are in terms of the symmetric Christoffel connection. 

In the next step, we generalize the action by considering the higher curvature defined in terms of the new Ricci scalar $\bar{R}$ 
\begin{equation}
  S_f=  \frac{1}{2\kappa^2}\int d^4x \sqrt{-g} f({\bar{R}})
  \label{actf}
\end{equation}

Generalized higher curvature action can be expressed as scalar tensor theory at an action level by introducing an auxiliary field. This mechanism is quite standard and widely used in the literature \cite{2010RvMP...82..451S,DeFelice:2010aj} where the scalar degree of freedom is the manifestation of higher curvature. Further, performing a conformal transformation of the form $g_{\mu\nu}\rightarrow\tilde{g}_{\mu\nu} = \Omega^2 {g}_{\mu\nu}$,  the action can be recast in Einstein frame with a scalar field and a potential which originates from higher curvature.  Following a similar technique for action (\ref{actf}), after a detailed calculation, Kumar et al\cite{2024EL....14729001K} arrived at the following action in Einstein frame
\begin{align}\label{actfe}
S_\mathrm{E} = \int d^4x \sqrt{-\tilde{g}} \Big[ 
& \frac{\tilde{R}}{2\kappa^2} 
 - \tfrac{1}{2}\tilde{g}^{ef}\partial_e \phi \partial_f \phi 
 - V(\phi) 
 - \tfrac{1}{2}\partial_{[e}Z_{ad]}\partial^{[e}Z^{ad]} \nonumber\\
& + \kappa \tilde{g}^{ac}\tilde{g}^{be}\tilde{g}^{fd}
   \big(\partial_{[e}\phi~Z_{ad]}\partial_{[f}Z_{bc]}\big) \nonumber\\
& - \tfrac{\kappa^2}{2}\tilde{g}^{ac}\tilde{g}^{be}\tilde{g}^{fd}
   \partial_{[e}\phi~Z_{ad]}\partial_{[f}\phi~Z_{bc]} \Big]~,
\end{align}

         where  $\tilde{g}_{\mu\nu}$ is the metric is the Einstein frame, $\tilde{R}$ is the usual Ricci scalar in Einstein frame, 
         $\Omega^2=f'=\frac{\partial f}{\partial\bar{R}}$, ~~ $\phi=\frac{2}{\kappa}\ln\Omega$ is the scalar field , ~~
         $V(\phi) = \frac{f'\bar{R} - f(\bar{R})}{2f'^2}$, ~ and ~
         $Z_{[ab]} \equiv \frac{\Omega^2}{\kappa}B_{[ab]}$, where $B_{[ab]}$ is a two rank antisymmetric field which appears in the solution of $K_{abc}$ as 
         $$K_{abc} = \left(\frac{2}{3\Omega^2}\right)\left\{\widetilde{g}_{ac} \partial_{b}\phi- \widetilde{g}_{bc} \partial_{a}\phi\right\} + \partial_{[b}B_{ac]}$$.
         
\section{Isotropic and Homogeneous universe}
We consider the background isotropic and homogeneous universe is described by the Friedmann-Robertson-Walker metric.
$$ds^2= -d{t}^2+\tilde{a}^2(t)(dx^2+dy^2+dz^2)$$ 
It is important to note that since both the scalar field and the antisymmetric tensor field are manifestations of curvature and the metric components depend only on time, both $\phi$ and $Z_{ad}$ will depend only on time.

 With the inclusion of torsion into the \( f(\bar{R}) \) theory, the action is modified in the Einstein frame, and three additional terms appear as a direct consequence of torsion in the spacetime geometry (\ref{actfe}). These extra terms represent the torsional contributions to the gravitational dynamics and modify the structure of the field equations. These three terms appear with three different coupling strengths. We consider the effects of these terms by including them systematically one by one according to their coupling strength. The first among these three has the kinetic terms of the torsion field, which has the same coupling strength as the scalar field.  In case 1, we will study the effect of this term only on the dynamics and ignore the terms with higher order of $\kappa$. In the action (\ref{actfe}), the second term associated with torsion has the next higher-order coupling in $\kappa$. In this term, the scalar field is coupled to the torsion field and its kinetic term. In case 2, we consider the action until the second torsion related term and investigate the consequences. Finally, in case 3, we consider the entire action with all torsion related terms.

\subsection{Case 1: }
In the first case, we include only the kinetic term of the torsion field as the contribution from the torsion tensor along with the scalar field in the action (\ref{actfe}), since they have the same coupling strength. We analyze its effect on the field equations and check the possibility of obtaining a bouncing solution.
\begin{align}\label{actfe1}
S_E = \int d^4x \sqrt{-\tilde{g}} \Bigg(
& \frac{\tilde{R}}{2\kappa^2} 
- \frac{1}{2} \tilde{g}^{\mu\nu} \partial_\mu \phi \, \partial_\nu \phi 
- V(\phi) \nonumber\\
& - \frac{1}{2} \partial_{[e} Z_{ad]} \, \partial^{[e} Z^{ad]} 
\Bigg)~,
\end{align}

The field equations governed by the above action (\ref{actfe1}) are
\begin{align}
        \tilde{R}_{\mu\nu} - \frac{1}{2} \tilde{g}_{\mu\nu} \tilde{R} = \kappa^2\left[ \partial_\mu \phi \partial_\nu \phi + 3\tilde{g}^{be} \tilde{g}^{fd}T_{e\mu d}T_{b\nu f} \right.\nonumber\\
        \left.-\tilde{g}_{\mu\nu} \left( \frac{1}{2} \tilde{g}^{ab} \partial_a \phi \partial_b \phi - V(\phi)- \frac{1}{2}T_{ead}T^{ead} \right)\right]
        \label{fldeq1}
\end{align}
\noindent
where \( T_{ead} = \partial_{[e} Z_{ad]} \) is the derivative of the antisymmetric tensor field $Z_{ad}$. \( T_{ead}\) is antisymmetric in all three indices. The scalar field follows the usual Klein Gordon equation 
\begin{equation}
\ddot{\phi} + 3\tilde{H}\dot{\phi} + \frac{\partial V}{\partial \phi} = 0.
\label{skg}
\end{equation}
while the torsion field follows
\begin{equation}
\partial_{\alpha} (\sqrt{-\tilde{g}}T^{\alpha\beta\gamma})=\partial_{\alpha} (\sqrt{-\tilde{g}}\partial^{[\alpha} Z^{\beta\gamma]})=0
\label{c1t}
\end{equation}
 With a FRW background, the diagonal field equations ($00$) and $(ii)$ look like
\begin{align}
3\tilde{H}^2 &= \kappa^2 \Bigg(
   \frac{\dot{\phi}^2}{2} + V 
   + \frac{1}{2} T_{ead} T^{ead} + 3 \tilde{g}^{be} \tilde{g}^{fd} T_{e0d} T_{b0f} 
\Bigg) ,
\label{c100}
\end{align}

\begin{align}
2\dot{\tilde{H}} + 3\tilde{H}^2 &= \kappa^2 \Bigg(
   -\frac{\dot{\phi}^2}{2} + V 
   + \frac{1}{2} T_{ead} T^{ead} - 3 \tilde{g}^{be} \tilde{g}^{fd} \tilde{g}^{ii} 
   T_{eid} T_{bif} 
\Bigg) ,
\label{c1ii}
\end{align}

On subtracting these equations, we get
\begin{equation}
\dot{\tilde{H}} =\frac{\kappa^2}{2}\left( -\dot{\phi}^2 - 3~\tilde{g}^{be} \tilde{g}^{fd} \tilde{g}^{11}~T_{e1d}T_{b1f}  - 3~\tilde{g}^{be} \tilde{g}^{fd}~T_{e0d}T_{b0f} \right)
\label{c1c1}
\end{equation}
The off diagonal field equations $(ij)$ in the FRW background are presented in Appendix 1. 
Due to the isotropy of the metric, from the off-diagonal field equations $(ij)$ we find that all components of \( T_{ead} \) are zero except for \( T_{123} \).  \( T_{123} \) can be found from equation (\ref{c1t}). Therefore, equation (\ref{c1c1}) simplifies to
\begin{equation}
\dot{\tilde{H}} =-\frac{\kappa^2}{2}\left(\dot{\phi}^2 + 3 ~ \tilde{g}^{11} \tilde{g}^{22} \tilde{g}^{33}~T_{123}^2 \right)
\end{equation}
Non-zero  \( T_{123} \) will influence the dynamics of the evolution, making it different from spacetime without torsion. However, a bouncing solution, which is characterized by a positive $\dot{H}$, is possible only if the terms within the bracket add up to a negative value. This eventually imply either one of the two fields or both fields (scalar and torsion) are phantom. So, a bounce cannot be realized though we consider both the scalar field and torsion field together.



\subsection{Case 2:}
In this case, we consider the next higher order term in $\kappa$ in the action (\ref{actfe}). This includes the term where the scalar field is coupled to the tensor field and its kinetic term. We study the impact of the coupled scalar and the torsion effect on the cosmological dynamics and whether this coupling supports a bouncing solution.
\begin{eqnarray}
\label{actfe2} 
&S_\mathrm{E}& = \int d^4x \sqrt{-\tilde{g}} \bigg[ \frac{\tilde{R}}{2\kappa^2} - \frac{1}{2} \tilde{g}^{ef} \partial_e \phi \partial_f \phi - V(\phi)  \nonumber\\ 
&-& \frac{1}{2} \partial_{[e} Z_{ad]} \partial^{[e} Z^{ad]}+\kappa \tilde{g}^{ac} \tilde{g}^{be} \tilde{g}^{fd} \left( \partial_{[e} \phi Z_{ad]} \partial_{[f} Z_{bc]} \right) \bigg].
\end{eqnarray}

The general field equations derived from the action (\ref{actfe2}) are
\begin{align}
\tilde{R}_{\mu\nu} - \tfrac{1}{2}\tilde{g}_{\mu\nu}\tilde{R} 
= \kappa^2 \Bigg[ 
& \partial_\mu \phi \, \partial_\nu \phi 
+ 3 \tilde{g}^{be} \tilde{g}^{fd} T_{e\mu d} T_{b\nu f} -\nonumber\\
&6\kappa \tilde{g}^{be} \tilde{g}^{fd} S_{e\mu d} T_{b\nu f} - \tilde{g}_{\mu\nu} \Big( 
   \tfrac{1}{2} \tilde{g}^{ef} \partial_e \phi \partial_f \phi 
   \nonumber\\
&  + V(\phi)  + \tfrac{1}{2} T_{ead} T^{ead}  - \kappa S_{ead} T^{ead} \Big) 
\Bigg] ,
\label{fldeq2}
\end{align}

Here, \(T_{ead} = \partial_{[e} Z_{ad]}\) and \(S_{ead} = \partial_{[e} \phi Z_{ad]}\). The field equations for the scalar field \(\phi\) and the antisymmetric tensor field \(Z_{\mu\nu}\) get modified due to the scalar-torsion coupling. 
\begin{eqnarray}
\ddot{\phi} + 3\tilde{H}\dot{\phi} + \frac{\partial V}{\partial \phi} + \frac{\kappa}{\sqrt{-\tilde{g}}} \partial_p(\sqrt{-\tilde{g}} Z_{ad} T^{pad}) = 0.\label{c2p}\\
\sqrt{-\tilde{g}}  \partial_e(\phi) T^{e\mu\nu} +\partial_e \left[ \sqrt{-\tilde{g}} (T^{e\mu\nu}-\kappa S^{e\mu\nu}  ) \right]=0.
\label{c2t}
\end{eqnarray}
  With a flat FRW background,  the diagonal Einstein field equations $(00)$ and ($ii$) look like
\begin{align}
3\tilde{H}^2 = \kappa^2 \Bigg[ 
& \tfrac{1}{2}\dot{\phi}^2 + V 
+ \tfrac{1}{2} T^{ead} \Big( T_{ead} - 2\kappa S_{ead} \Big)\nonumber\\
&+ 3 \tilde{g}^{be} \tilde{g}^{fd} 
   T_{b0f} \Big( T_{e0d} - 2\kappa S_{e0d} \Big) 
\Bigg] ,
\label{c200}
\end{align}

\begin{align}
2\dot{\tilde{H}} + 3\tilde{H}^2 = \kappa^2 \Bigg[ 
& -\tfrac{1}{2}\dot{\phi}^2 + V 
+ \tfrac{1}{2} T^{ead} \Big( T_{ead} - 2\kappa S_{ead} \Big) \nonumber\\
& - 3 \tilde{g}^{ii}\tilde{g}^{be}\tilde{g}^{fd} \,
   T_{bif} \Big( T_{eid} - 2\kappa S_{e1d} \Big) 
\Bigg] ,
\label{c2ii}
\end{align}

The off diagonal field equations $(ij)$ are presented in the Appendix 1. From the analysis of the off-diagonal field equations $(ij)$, a crucial relationship emerges between the antisymmetric tensor field \(T_{ead}\) and the scalar-torsion coupling term \(S_{ead}\) 
\begin{equation}
T_{ead} = 2\kappa S_{ead}.
\label{c2}
\end{equation}

\noindent
Since both $\phi$ and $Z_{ad}$ depend on time, the relations (\ref{c2}) imply that only the components \(Z_{12}\), \(Z_{23}\), and \(Z_{31}\) remain non-zero and are related to the scalar field \(\phi\) as:
\begin{equation}
Z_{12} = Z_{23} = Z_{31} = \exp(2\kappa\phi).
\label{c2z}
\end{equation}
 Substituting the relation (\ref{c2}) into equation (\ref{c2p}) simplifies the equation for \(\phi\), resulting in the standard Klein-Gordon equation:
\begin{equation}
\ddot{\phi} + 3\tilde{H}\dot{\phi} + \frac{\partial V}{\partial \phi} = 0.
\end{equation}

Using (\ref{c2}) into the original field equations (\ref{c200}) and (\ref{c2ii}) simplifies them, leading to
\begin{eqnarray}   
3\dot{\tilde{H}}^2=\kappa^2\left( \frac{\dot{\phi}^2}{2} + V\right)\nonumber\\
2\dot{\tilde{H}}+3\dot{\tilde{H}}^2=\kappa^2\left( -\frac{\dot{\phi}^2}{2} + V\right)
\label{frwg}
\end{eqnarray}
These equations are same as the ones in Einstein gravity with minimally coupled scalar field. Thus leading to the same conclusion as in Einstein gravity. This shows that even though we consider scalar-torsion coupling in isotropic homogeneous background, it is not possible to find a bouncing cosmological solution.


\subsection{Case 3: }
 Finally, we consider the complete action (\ref{actfe}) with all three torsion contributions and explore whether this full torsional extension of \( f(\bar{R}) \) gravity can support a bouncing solution in an isotropic and homogeneous universe without violating the energy conditions.
\begin{align}
S_\mathrm{E} = & \int d^4x \sqrt{-\tilde{g}} \Bigg[
\frac{\tilde{R}}{2\kappa^2} 
- \tfrac{1}{2} \tilde{g}^{ef} \partial_e \phi \, \partial_f \phi
- V(\phi) 
 \nonumber\\
& - \tfrac{1}{2} \partial_{[e} Z_{ad]} \, \partial^{[e} Z^{ad]} + \kappa \tilde{g}^{ac} \tilde{g}^{be} \tilde{g}^{fd} 
   \Big( \partial_{[e} \phi \, Z_{ad]} \, \partial_{[f} Z_{bc]} \Big) \nonumber\\
& - \tfrac{\kappa^2}{2} \tilde{g}^{ac} \tilde{g}^{be} \tilde{g}^{fd} 
   \Big( \partial_{[e} \phi \, Z_{ad]} \, \partial^{[e} \phi \, Z^{ad]} \Big)
\Bigg]~.
\label{actfe3}
\end{align}

The field equations corresponding to this action are
\begin{align}
\tilde{R}_{\mu\nu} - \tfrac{1}{2}\tilde{g}_{\mu\nu}\tilde{R} 
= \kappa^2 \Bigg[
& \partial_\mu \phi \, \partial_\nu \phi 
+ 3 \tilde{g}^{be} \tilde{g}^{fd} \, T_{e\mu d} T_{b\nu f} \nonumber\\
& - 6 \tilde{g}^{be} \, S_{e\mu d} T_{b\nu f} 
+ 3 \tilde{g}^{be} \tilde{g}^{fd} \, S_{e\mu d} S_{b\nu f} \nonumber\\
& - \tilde{g}_{\mu\nu} \Big(
   \tfrac{1}{2}\tilde{g}^{ef} \partial_e \phi \, \partial_f \phi 
   + V + \tfrac{1}{2} T_{ead}T^{ead} \nonumber\\
&    - S_{ead}T^{ead} 
   + \tfrac{1}{2} S_{ead}S^{ead} 
   \Big)
\Bigg] ,
\label{fldeq3}
\end{align}
The field equations for the scalar field \(\phi\) and the antisymmetric tensor field \(Z_{\mu\nu}\) also get modified due to the inclusion of the last term in the action (\ref{actfe3})
\begin{align}
\ddot{\phi} + 3\tilde{H}\dot{\phi} + \frac{\partial V}{\partial \phi} 
&+ \frac{\kappa}{\sqrt{-\tilde{g}}} 
   \partial_p \bigg[ \sqrt{\tilde{g}} \, Z_{ad} 
   (T^{pad}-\kappa S^{pad}) \bigg] = 0,
\label{c3p} \\[6pt]
\kappa \sqrt{-\tilde{g}} \, \partial_e(\phi) 
&\bigg[ T^{e\mu\nu}-\kappa S^{e\mu\nu}\bigg] 
+ \partial_e \bigg[ \sqrt{-\tilde{g}} 
   (T^{e\mu\nu}-\kappa S^{e\mu\nu}) \bigg] = 0.
\label{c3t}
\end{align}

Combining equations (\ref{c3p}) and (\ref{c3t}) one can get
\begin{equation}
\ddot{\phi} + 3\tilde{H}\dot{\phi} + \frac{\partial V}{\partial \phi} -\kappa (T^{pad}-\kappa S^{pad}) (\partial_p Z_{ad} -\kappa\partial_p \phi Z_{ad}) = 0.
\label{c3sc}
\end{equation}
For a flat isotropic homogeneous background, the diagonal equations $(00)$ and $(ii)$ look like
\begin{align}
3\tilde{H}^2 &= \kappa^2 \Bigg[
   \frac{\dot{\phi}^2}{2} + V 
   + \tfrac{1}{2} \big( T_{ead} T^{ead} - 2 S_{ead} T^{ead} +
    \nonumber\\
&\ S_{ead} S^{ead}\big)+ 3~\tilde{g}^{be} \tilde{g}^{fd} 
   \big( T_{e0d} T_{b0f} - 2 S_{e0d} T_{b0f}  \nonumber\\
 &\  + S_{e0d} S_{b0f} \big)
   \Bigg],
\label{c300} \\[6pt]
2\dot{\tilde{H}} + 3\tilde{H}^2 &= \kappa^2 \Bigg[
   -\tfrac{\dot{\phi}^2}{2} + V 
   + \tfrac{1}{2} \big( T_{ead} T^{ead} - 2 S_{ead} T^{ead}  \big) \nonumber\\
&\quad + S_{ead} S^{ead} - 3~\tilde{g}^{ii} \tilde{g}^{be} \tilde{g}^{fd} 
   \big( T_{eid} T_{bif} - \nonumber\\
   &\quad 2 S_{eid} T_{bif} 
   + S_{eid} S_{bif} \big)
   \Bigg].
\label{c3ii}
\end{align}
The off-diagonal field equations are given in Appendix 1.
These equations  impose the following constraints
\begin{equation}
 T_{ead}T_{ead}-2\kappa T_{ead}S_{ead} +\kappa^2 S_{ead}S_{ead}=0.
 \label{c3}
\end{equation}
Using equation~(\ref{c3}), the field equations again reduce to the form of (\ref{frwg}) 
\begin{equation}
\dot{\tilde{H}} =-\frac{\kappa^2}{2}{\dot{\phi}^2}
\label{frwc}
\end{equation}
Thus by imposing the constraint equations obtained from off-diagonal equations in both cases 2 and 3, the field equations become similar to the one obtained in Einstein gravity with minimally coupled scalar field. Since using (\ref{c3}) the field equations reduces to the form (\ref{frwg}) the scalar field equation should follow the Klein Gordon equation (\ref{skg}). For such consistency, the last term in (\ref{c3sc}) should be zero, giving $Z_{ij}=\exp({\kappa\phi})$, a relation similar to (\ref{c2z}). Interestingly in both the cases, 
the antisymmetric torsion field arising out of antisymmetric nature of curvature gets identified with a function of the scalar degree of freedom which is the manifestation of the higher curvature. In the different context , Gonzalez Quaglia et al have also found that torsion, in scalar-tensor models are fully determined by the scalar field \cite{2023EPJP..138...93G}.

Equation (\ref{frwc}) clearly dismiss any chance of obtaining a bouncing cosmological solution without violation of strong energy condition. Thus, the inclusion of torsion terms does not  alter the fundamental structure of the equations as well as the chance of obtaining a bouncing solution in an isotropic and homogeneous universe in Einstein frame.

Thus, we conclude that, in an isotropic and homogeneous universe, a bounce solution is not feasible in the Einstein frame without violating energy conditions. Isotropy and homogeneity impose some stringent conditions on torsion scalar coupling, which in turn express the torsion field in terms of the scalar field. Driven by this observation, we now relax the condition of isotropy and homogeneity to explore alternative possibilities of bounce. This approach aims to expand our understanding of the dynamics of the universe beyond the standard assumptions of isotropy and homogeneity.

\section{Inhomogeneous and Anisotropic Universe}

We consider an anisotropic and inhomogeneous metric such that we deviate very little from the standard assumption of the universe.  
\begin{equation}  
ds^2 = -d{t}^2 + \tilde{a}^2({t})e^{-2d~({x}+{y}+{z})} \left( d{x}^2 + d{y}^2 + d{z}^2 \right),  
\label{aim}
\end{equation}  
where \(d\) is a very small constant. This metric introduces very 
 small anisotropy and inhomogeneity  in the early universe, allowing us to study the dynamics while  retrieving isotropy and homogeneity in the limit $d\rightarrow very small.$ We will further denote $\frac{\dot{\tilde{a}}}{\tilde{a}}$ as $\tilde{H}$ and $\left[\frac{\ddot{\tilde{a}}}{\tilde{a}}-\left(\frac{\dot{\tilde{a}}}{\tilde{a}}\right)^2\right]$ as $\dot{\tilde{H}}$.

Since the above metric depend on both space and time and also both the scalar field and the antisymmetric tensor field are manifestations of curvature, both $\phi$ and $Z_{ad}$ will now depend on space as well as time. As discussed earlier, the torsion-dependent terms in action (\ref{actfe}) appear with different coupling strengths of $\kappa$, we will include them one by one in the increasing order of $\kappa$, like the isotropic and homogeneous case.

\subsection{Case 1:}
In the anisotropic and inhomogeneous background we first consider the action (\ref{actfe1}). As stated earlier, this action is a truncated version of the full action (\ref{actfe}) where only the terms of same coupling strength have been considered. The general field equations are given by (\ref{fldeq1}) and the scalar field and torsion field follows equation (\ref{skg}) and (\ref{c1t}) respectively. The diagonal field equations in this background look like
\begin{equation}
\label{e00}
\begin{split}
(00)\;\rightarrow\;\frac{1}{\kappa^2}\big(3\tilde{H}^2 - 3d^2 \tilde{g}^{11}\big) 
&= \tfrac{1}{2}\Big( (\partial_0 \phi)^2 
   + \tilde{g}^{11} (\partial_1 \phi)^2+ \tilde{g}^{22} (\partial_2 \phi)^2  \\
&\quad 
   + \tilde{g}^{33} (\partial_3 \phi)^2 \Big)+ V + \tfrac{1}{2} T^{ead} T_{ead}  \\
&\quad + 3 \tilde{g}^{be}\tilde{g}^{fd}\tilde{g}^{00} T_{b0f}T_{e0d},
\end{split}
\end{equation}

\begin{equation}
\label{e11}
\begin{split}
(11)\;\rightarrow\;\frac{1}{\kappa^2}\big(2\dot{\tilde{H}} + 3\tilde{H}^2 - d^2 \tilde{g}^{11}\big) 
&= \tfrac{1}{2}\Big( -(\partial_0 \phi)^2 
   - \tilde{g}^{11} (\partial_1 \phi)^2  \\
&\quad + \tilde{g}^{22} (\partial_2 \phi)^2  + \tilde{g}^{33} (\partial_3 \phi)^2 \Big)   \\
&\quad + V  + \tfrac{1}{2} T^{ead}T_{ead}\\
&- 3 \tilde{g}^{be}\tilde{g}^{fd}\tilde{g}^{11} T_{b1f}T_{e1d},
\end{split}
\end{equation}

\begin{equation}
\label{e22}
\begin{split}
(22)\;\rightarrow\;\frac{1}{\kappa^2}\big(2\dot{\tilde{H}} + 3\tilde{H}^2 - d^2 \tilde{g}^{22}\big) 
&= \tfrac{1}{2}\Big( -(\partial_0 \phi)^2 
   + \tilde{g}^{11} (\partial_1 \phi)^2 \\
&- \tilde{g}^{22} (\partial_2 \phi)^2 
   + \tilde{g}^{33} (\partial_3 \phi)^2 \Big) \\
& + V + \tfrac{1}{2} T^{ead}T_{ead} \\
 &  - 3 \tilde{g}^{be}\tilde{g}^{fd}\tilde{g}^{22} T_{b2f}T_{e2d},
\end{split}
\end{equation}

\begin{equation}
\label{e33}
\begin{split}
(33)\;\rightarrow\;\frac{1}{\kappa^2}\big(2\dot{\tilde{H}} + 3\tilde{H}^2 - d^2 \tilde{g}^{33}\big) 
&= \tfrac{1}{2}\Big( -(\partial_0 \phi)^2 
   + \tilde{g}^{11} (\partial_1 \phi)^2 \\
& + \tilde{g}^{22} (\partial_2 \phi)^2 
   - \tilde{g}^{33} (\partial_3 \phi)^2 \Big) \\
& + V + \tfrac{1}{2} T^{ead}T_{ead} \\
  & - 3 \tilde{g}^{be}\tilde{g}^{fd}\tilde{g}^{33} T_{b3f}T_{e3d}.
\end{split}
\end{equation}
The off diagonal field equations are listed in Appendix 2. The off-diagonal equations impose crucial constraints on the scalar field and the torsion fields. The equations dictate the relation 
\begin{equation}   
\partial_1{\phi} = \partial_2{\phi} = \partial_3{\phi}
\end{equation}
which implies $\phi$ is a $f(x+y+z)$. The ratio between different torsion components follow the relations
\begin{equation}\label{a2tc1}
\frac{T_{012}}{T_{023}} =  \frac{T_{012}}{T_{031}}= \frac{T_{031}}{T_{023}}=1
\end{equation}
suggesting $T_{012}=T_{013}=T_{023}$. $T_{123}$ is related to other components by (\ref{a2tcc}). 
Finally combining the conditions with the equations (\ref{e00})-(\ref{e33}), the $\dot{\tilde{H}}$ equation can be  written as 
\begin{equation}
\begin{split}
\dot{\tilde{H}} &= -\kappa^2\Bigg(
    \tfrac{1}{2}\big(\dot{\phi}^2 + \tilde{g}^{11}(\partial_1\phi)^2\big) \\
&+ 3\big(T^{012}T_{012} + T^{123}T_{123}\big)
\Bigg)- d^2\,\tilde{g}^{11}\,.
\end{split}
\label{hdot_split}
\end{equation}
In the right hand side, all the terms are quadratic, hence positive. So the conclusion remain same as in earlier cases.
Thus even if the condition of symmetry in spacetime is relaxed, a bouncing solution is not favoured, if only the first torsion term is considered. 

\subsection{Case 2: }

We now consider the action (\ref{actfe2}) where  the coupled scalar-tensor term is included which has the next order of coupling of $\kappa$. The general field equations and the equations for the scalar and the torsion field for (\ref{actfe2}) are given by (\ref{fldeq2}), (\ref{c2p}) and (\ref{c2t}) respectively. In the anisotropic and inhomogeneous background (\ref{aim}), the diagonal field equations ($00$) and $(ii)$ look like
\begin{equation}
\label{a00}
\begin{split}
(00)\;\rightarrow\;\frac{1}{\kappa^2}\big(3\tilde{H}^2 &- 3d^2 \tilde{g}^{11}\big) 
= \tfrac{1}{2}\Big( \dot{\phi}^2 + \tilde{g}^{11}(\partial_1\phi)^2 \\
&+ \tilde{g}^{22}(\partial_2\phi)^2 + \tilde{g}^{33}(\partial_3\phi)^2 \Big) \\
&+ V + \tfrac{1}{2} T^{ead}\big( T_{ead} - 2\kappa S_{ead} \big) \\
&\quad + 3\,\tilde{g}^{be}\tilde{g}^{fd}\tilde{g}^{00}\,
    T_{b0f}\big( T_{e0d} - 2\kappa S_{e0d} \big)\,,
\end{split}
\end{equation}

\begin{equation}
\label{a11}
\begin{split}
(11)\;\rightarrow\;\frac{1}{\kappa^2}\big(2\dot{\tilde{H}} +& 3\tilde{H}^2 - d^2 \tilde{g}^{11}\big) 
= \tfrac{1}{2}\Big( -\dot{\phi}^2 - \tilde{g}^{11}(\partial_1\phi)^2 \\
&\quad + \tilde{g}^{22}(\partial_2\phi)^2 + \tilde{g}^{33}(\partial_3\phi)^2 \Big) \\
&\quad + V + \tfrac{1}{2} T^{ead}\big( T_{ead} - 2\kappa S_{ead} \big) \\
&\quad - 3\,\tilde{g}^{be}\tilde{g}^{fd}\tilde{g}^{11}\,
    T_{b1f}\big( T_{e1d} - 2\kappa S_{e1d} \big)\,,
\end{split}
\end{equation}

\begin{equation}
\label{a22}
\begin{split}
(22)\;\rightarrow\;\frac{1}{\kappa^2}\big(2\dot{\tilde{H}} + & 3\tilde{H}^2 - d^2 \tilde{g}^{22}\big) 
= \tfrac{1}{2}\Big( -\dot{\phi}^2 + \tilde{g}^{11}(\partial_1\phi)^2 \\
&\quad - \tilde{g}^{22}(\partial_2\phi)^2 + \tilde{g}^{33}(\partial_3\phi)^2 \Big) \\
&\quad + V + \tfrac{1}{2} T^{ead}\big( T_{ead} - 2\kappa S_{ead} \big) \\
&\quad - 3\,\tilde{g}^{be}\tilde{g}^{fd}\tilde{g}^{22}\,
    T_{b2f}\big( T_{e2d} - 2\kappa S_{e2d} \big)\,,
\end{split}
\end{equation}

\begin{equation}
\label{a33}
\begin{split}
(33)\;\rightarrow\;\frac{1}{\kappa^2}\big(2\dot{\tilde{H}} + & 3\tilde{H}^2 - d^2 \tilde{g}^{33}\big) 
= \tfrac{1}{2}\Big( -\dot{\phi}^2 + \tilde{g}^{11}(\partial_1\phi)^2 \\
&\quad + \tilde{g}^{22}(\partial_2\phi)^2 - \tilde{g}^{33}(\partial_3\phi)^2 \Big) \\
&\quad + V + \tfrac{1}{2} T^{ead}\big( T_{ead} - 2\kappa S_{ead} \big) \\
&\quad - 3\,\tilde{g}^{be}\tilde{g}^{fd}\tilde{g}^{33}\,
    T_{b3f}\big( T_{e3d} - 2\kappa S_{e3d} \big)\,.
\end{split}
\end{equation}
The off-diagonal equations are listed in Appendix 2.
The constraint relations in $\phi$ obtained from these off-diagonal equations imply that the spatial dependence of $\phi$ is of the form $f(x+y+z)$ like in the previous case.  In addition, the constraint relations in $T_{ead}$ and $S_{ead}$ obtained from the off-diagonal equations suggest $T_{012}=T_{013}=T_{023}$ and $S_{012}=S_{013}=S_{023}$. The relationship involving the $T_{123}$ with the other $T$ components  and $S_{123}$ with the other components $S$ are given by (\ref{a2tc}) and (\ref{a2sc}) respectively.

After using these relations in equations (\ref{a00})-(\ref{a22}) and (\ref{a33}) and further simplifying we get 
\begin{equation}
2 {\tilde{\dot{H}}} = \frac{\kappa^2 d^2\dot{\phi}^2}{\kappa^2(\partial_1 \phi)^2 - {d^2} }- 3d^2 g^{11}.
\label{a2}
\end{equation}
So, for the first time we find a scenario, where a positive $\dot{\tilde{H}}$ can be achieved at a particular epoch without a phantom field and satisfy the minimum condition for a bounce. 

Now we look for the bouncing solution as the minimum requirement is met. Here, after taking into account all the symmetries and constraint relations between different field components, the system has fewer field equations than dynamical quantities to determine. So,  we consider a form for $\tilde{a}$ suitable for bounce.\\
\begin{itemize}
\item 
{\bf Form of the scale factor:} \\
Different forms of bouncing scale factors are available in the literature\cite{2020NuPhB.95915159O}. Among them, scale factor of the form $\tilde{a}=a_0(1+\chi  t^2)^{n},$ is widely used\cite{2022Univ....8..636L,2024EPJC...84.1202T,2025arXiv250320709M}. Not only, this form satisfies the conditions required for a bouncing scenario in section 2, it has been used in different kind of investigations of early universe. In a bottoms up  approach Odintsov et al\cite{2020NuPhB.95915159O} related observational indices $(n_s,r)$ in inflationary scenario to the the slow roll parameters and reconstructed different scalar factor via conformal transformation. With a similar form, Karimzadeh et al.\cite{2019arXiv190204406K} showed  the possibility of achieving effective phantom behavior without phantom fields in modified teleparallel gravity. Here we consider
\begin{equation}
    \tilde{a}=a_i(a_b+At^2)^{1/n},
    \label{sc}
\end{equation}
where $a_i$ is the scale factor at the bouncing epoch at t = 0. For the sake of brevity, we take $a_i = 1$. And we have taken $n=3$.
The expressions for the Hubble parameter and the deceleration parameter are
\begin{equation}
\tilde{H}=\frac{2At}{3(a_b+At^2)}~{\text{and}}~q=\frac{1}{2}-\frac{3}{2At^2}
\label{hnq}
\end{equation}
We plot both the parameters below in fig (\ref{fighq}) for $a_b=1$ and $A=0.05$. We used these values of $a_b$ and $A$ in our further calculations. The plots are similar for different values of $A$. The Hubble parameter is negative before the bouncing epoch, vanishes at the time of bounce and is positive after bounce as is required for a bouncing scenario. The deceleration parameter diverge at bounce, but is symmetric and negative on both side of bounce in the early time.
\begin{figure}[h!]
    \centering
    \caption{Plots of the Hubble parameter $H$ and deceleration parameter $q$ for $n=3$ and $A=0.05$.}
    \label{fighq}
    \includegraphics[height=3.5cm, width=0.48\linewidth]{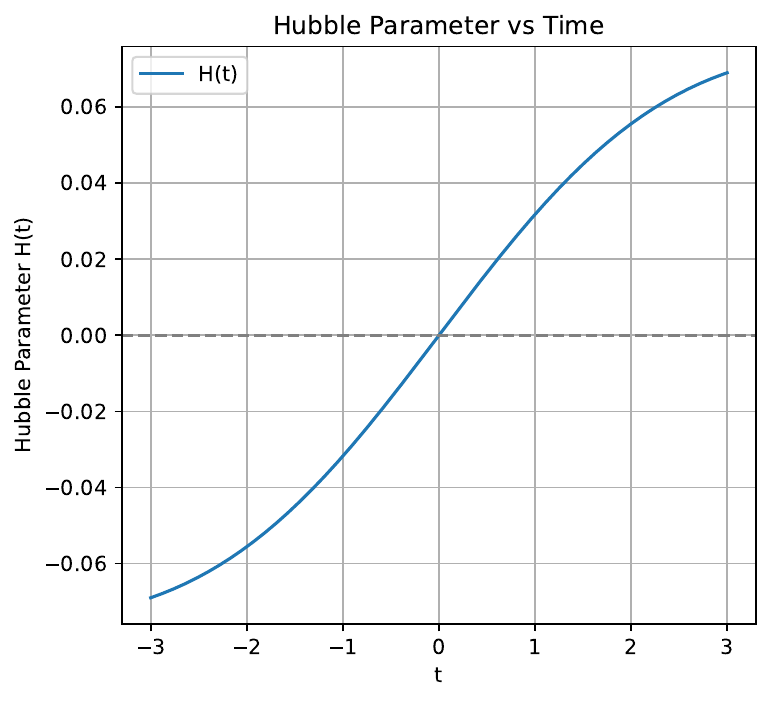}
    \includegraphics[height=3.5cm, width=0.48\linewidth]{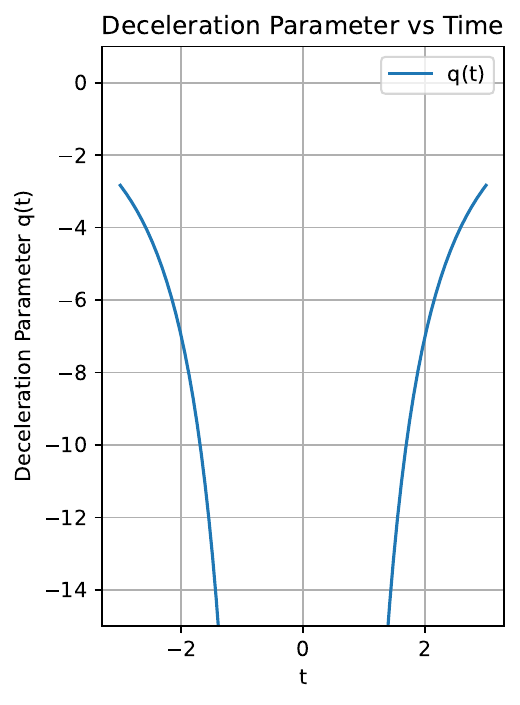}
\end{figure}

\item
{\bf Solution for the Scalar field:}\\
We consider the ratio given in (\ref{a2tc}) to be a constant 
$\alpha$ i.e  
\begin{equation}
    \frac{T_{123}}{T_{023}}=\frac{T_{123}}{T_{012}}=\frac{T_{123}}{T_{013}}=\alpha
    \label{a2tc1}
\end{equation}
and thus
\begin{equation}
\partial_0{\phi} = \alpha \tilde{g}^{11}\frac{\left[\kappa^2 (\partial_1{\phi})^2-d^2 \right]}{\kappa^2\partial_1{\phi}} 
\label{a2p01}
\end{equation}
Substituting this expression for \(\partial_0{\phi}\) into the equation (\ref{a2}) and integrating, we obtain the solution for \(\phi\):  
\begin{equation}
\begin{split}
\phi =&\frac{1}{2\kappa^2}\ln\Bigg|\frac{\alpha~d~e^{2d(x+y+z)}}{\tilde{a}^2} - \frac{3d}{2\alpha} + \\
&\sqrt{\left(\frac{\alpha d e^{2d(x+y+z)}}{\tilde{a}^2} - \frac{3d}{2\alpha}\right)^2 - 2\left[\frac{4A(1-At^2)}{3(1+At^2)^{2}} + \left(\frac{3d}{2\alpha}\right)^2\right]} \Bigg|+ C 
\end{split}
\label{a2s}
\end{equation}
where \(C\) is an integration constant.
Substituting $\phi$ as obtained in (\ref{a2s}) and the scale factor in the equation (\ref{sc}) one can obtain the expression for the potential 
\begin{equation}
\begin{split}
   V = & \;\frac{4A(1-At^2)}{3(1+At^2)^{2}} 
        + \frac{12A^2t^2}{9(1+At^2)^2} \\
      & - 6d^2(1+At^2)^{-2/3} \exp\!\big(2d(x+y+z)\big) - 1 \\
      & + \frac{1}{\alpha^2} \cdot 
        \frac{(1 + At^2)^{4/3}}{\exp\!\big(4d(x+y+z)\big)} 
        \cdot \Big[
           \frac{4A(1 - 3At^2)}{3(1 + At^2)^2}\\
          & + \frac{3d^2 \exp\!\big(2d(x+y+z)\big)}{(1 + At^2)^{2/3}}
        \Big]
\end{split}
\label{a2V}
\end{equation}

In the fig (\ref{figpv}) we plot $\phi$ and the potential $V$. $\phi$ depends on both space and time. To understand the profile of $\phi$, we vary time from (-3 to +3) and all the three space coordinates from (-1 to +1). We have presented a 3D plot of $\phi$   with respect to $t$ and $x$. With respect to other space coordinates also, it remains the same. Since the profile of $\phi$ remains same for  the space coordinates, we also plot $\phi$ with respect to $t$ only for some value of the spatial parameters $x,y,z=0.1$. 
Both the plot show $\phi$ has a maxima at the bouncing epoch. We have also plotted the potential with respect to time and $\phi$. V also have a maxima at $t=0$. The flow of time in $V$ vs $\phi$ plot is indicated with arrow where the maxima occurs at maximum of $\phi$ at $t=0$
\begin{figure}[h!]
    \centering
    \caption{1st row - Plots of $\phi$ in 3 dimension and 2 dimension. 2nd row -Plots of the potential with respect to time and $\phi$. The values $\alpha = 1000$, $C = 8$, $\kappa$ and $d = 0.008$ were considered to plot these graphs. }
    \includegraphics[height=3.5cm, width=0.48\linewidth]{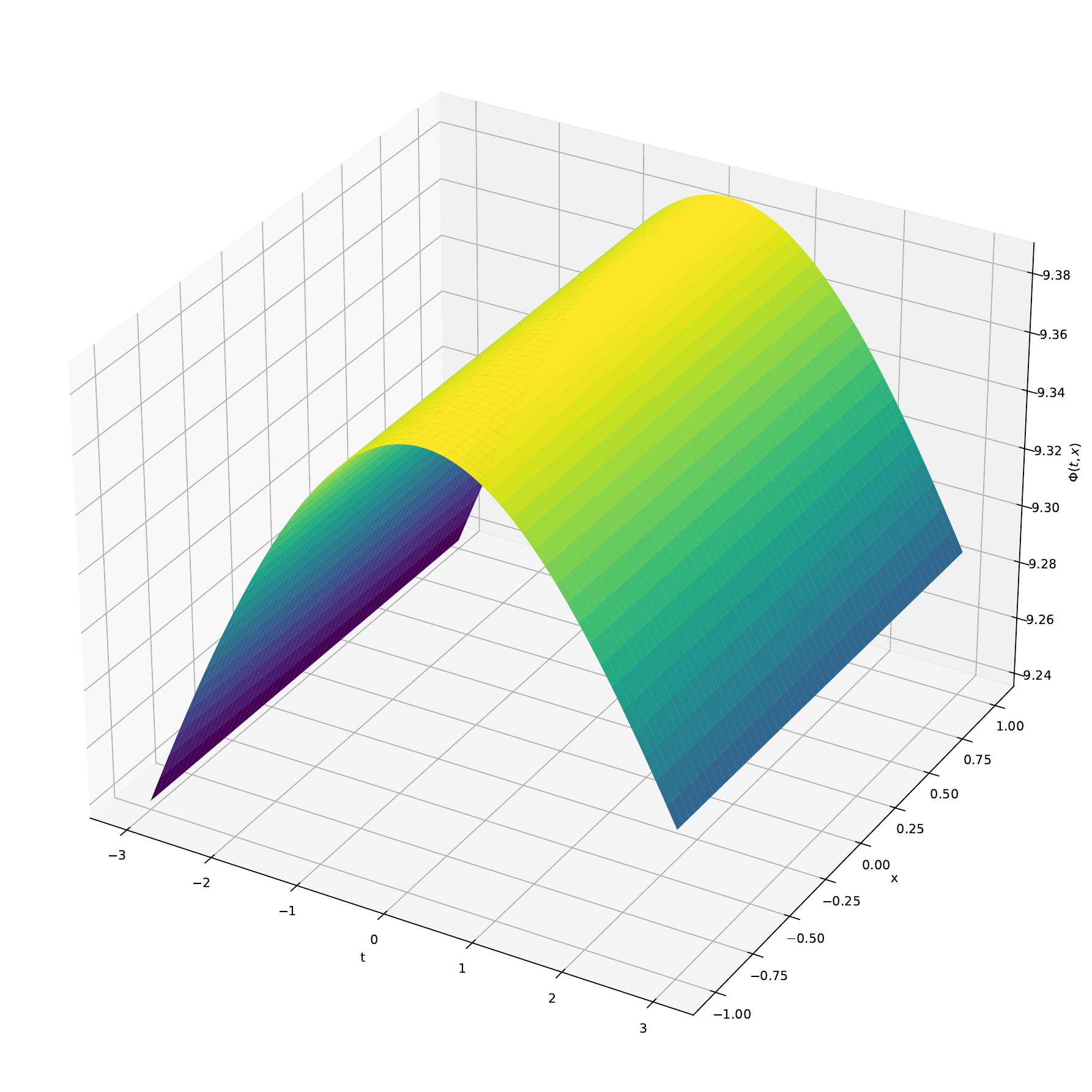}
    \includegraphics[height=3.5cm, width=0.48\linewidth]{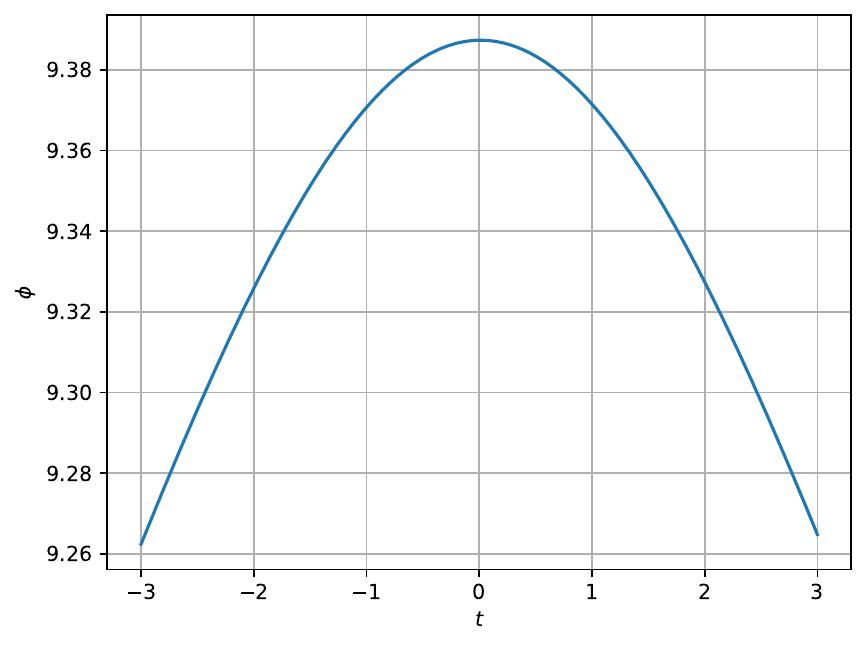}\\
    \includegraphics[height=3.5cm, width=0.48\linewidth]{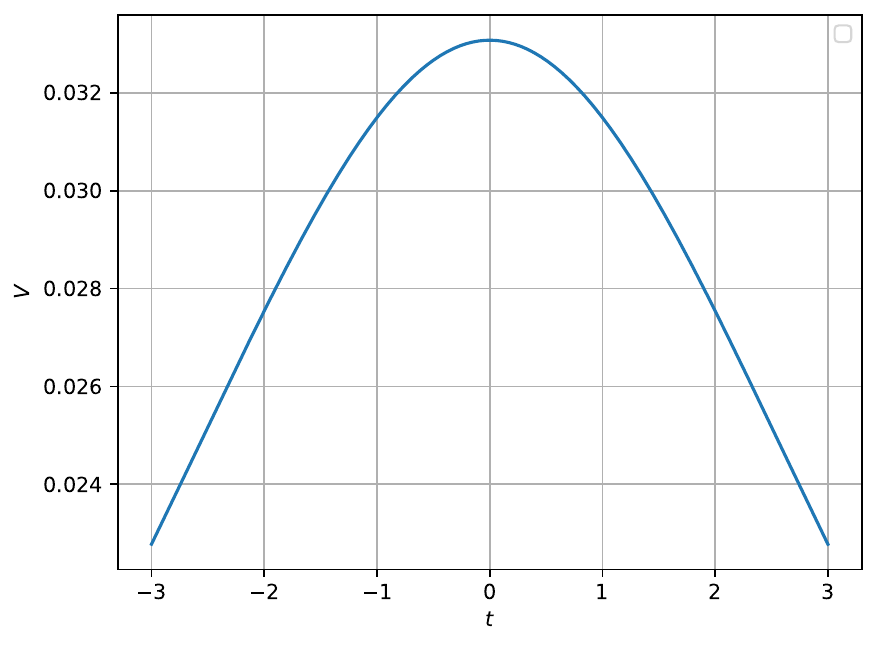}
    \includegraphics[height=3.5cm, width=0.48\linewidth]{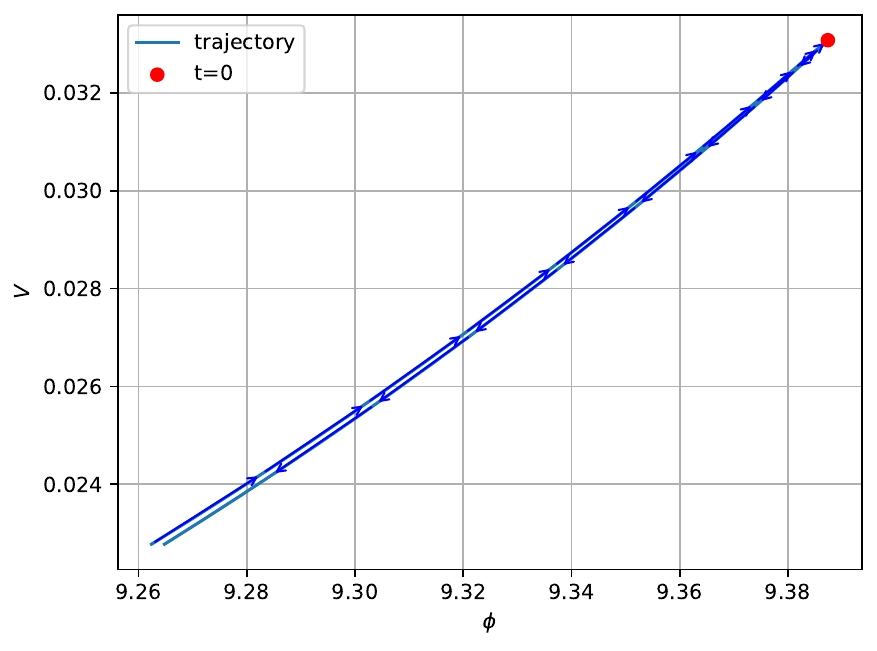}
    \label{figpv}
\end{figure}
\item
{\bf Solution for the torsion fields}

Finally we look for the solution of torsion field. From the symmetry of the equations, metric tensor and solution of $\phi$, it can be understood that the spatial dependence of the different component of the torsion field is a $f(x+y+z)$. Guided by the equations and earlier solutions we consider the spatial dependence to be  $\exp{(x+y+z)}$. Thus, we consider the following forms 

\begin{equation}
\begin{aligned}
Z_{01} &= f_{1}(t)\exp(-d(x+y+z))\\
Z_{02} &= f_2(t) \exp(-d(x+y+z))\\
Z_{03} &= f_3(t) \exp(-d(x+y+z))\\
Z_{12} &= P_{1}(t) \exp(-d(x+y+z)) \\
Z_{23} &= P_2(t) \exp(-d(x+y+z)) \\
Z_{31} &= P_3(t) \exp(-d(x+y+z))
\end{aligned}
\label{zs}
\end{equation}
From the off-diagonal field equations, we can have relation between $S$ components and $T$ components, like 
\begin{equation}
\frac{S_{123}}{S_{012}}=\frac{T_{123}}{T_{012}} = \alpha
\label{a2ts}
\end{equation}
Using the forms of torsion tensor given in (\ref{zs}) and in equation (\ref{a2ts}), we obtain two relations:
\begin{eqnarray}
d(P_1 + P_2 + P_3) = \alpha\left[d(f_1 - f_2)-\dot{P}_1 \right],
\label{a2pf1}\\
\partial_1 \phi (P_1 + P_2 + P_3) = \alpha \bigg[(f_1 - f_2) \partial_1 \phi + P_1  \partial_0{\phi} \bigg].
\label{a2pf2}
\end{eqnarray}
The relation (\ref{a2tc1}) gives another set of relations
\begin{eqnarray}
\label{pf}
  \frac{1}{d}\{\dot{P_1} - \dot{P_2}\} &= f_1 + f_3 - 2f_2 \nonumber\\
  \frac{1}{d}\{\dot{P_2} - \dot{P_3}\} &= f_1 + f_2 - 2f_3 \\
  \frac{1}{d}\{\dot{P_1} - \dot{P_2}\} &= f_2 + f_3 - 2f_1\nonumber
\end{eqnarray}
From equations (\ref{a2pf1}) and (\ref{a2pf2}), we obtain the following relation:
\begin{equation}
\frac{\dot{P}_1}{P_1} = -d\frac{\partial_0{\phi}}{\partial_1 \phi}.
\label{a2p1c}
\end{equation}
Similar relations can be obtained for $P_2$ and $P_3$ and hence 
\begin{equation}
\label{pp}
\frac{\dot{P}_1}{P_1} = \frac{\dot{P}_2}{P_2} = \frac{\dot{P}_3}{P_3}= -d\frac{\partial_0{\phi}}{\partial_1 \phi}.
\end{equation}
which prompts $P_1=P_2=P_3=P$. Following this relation between the $P$'s, equation (\ref{pf}) dictates $f_1=f_2=f_3$. \\
Substituting the solution of $\phi$ readily gives the solution for $P$ to be
\begin{equation}
P = \exp\left( -\frac{4A}{3\alpha} t \left[1 - \frac{7A t^2}{9} \right] \frac{e^{-2d(x + y + z)}}{d^2} - \frac{3t}{\alpha} + C_1 \right)
\label{a2p1}
\end{equation}

\noindent
where \(C_1\) is a constant. Since, none of the off-diagonal relations involves any time derivative of $f$'s, hence we can choose $f_1=f_2=f_3=\text{constant}~C_2$. So, $Z_{01}=Z_{02}=Z_{03}$ and they depend only on space, while the other 3 are also equal $Z_{12}=Z_{23}=Z_{31}$, but depend on space and time both. In figure (\ref{figz}), we present $Z_{12}$ as it varies with time. We have taken same values for the constants as earlier with $C_1 =0$ and $x,y,z=0.1$
\begin{figure}[h!]
    \centering
    \caption{Plot of $Z_{12}$ with respect to time}
     \includegraphics[height=5cm, width=0.8\linewidth]{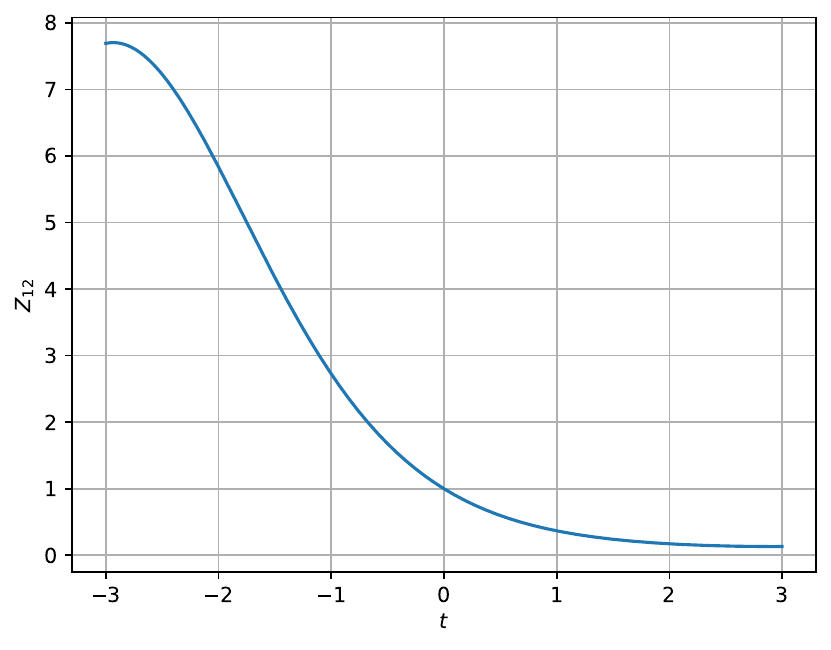}
    \label{figz}
\end{figure}
Interestingly, the torsion fields decays in time. This behaviour is consistent with the observed absence of the torsion fields in the universe.
Thus, we finally find an exact solution representing a bounce in an anisotropic and inhomogeneous  background. We also find the analytical solutions of the scalar field $\phi$ and the torsion field $Z_{\mu\nu}$ which are the manifestation of the higher curvature and antisymmetric affine connection respectively. 
\item 
{\bf Reconstruction of $f({\bar R})$ :}\\
Our entire analysis has been carried out in the Einstein frame. We  started with a generalized Einstein Cartan theory or $f(\bar{R})$. It is quite possible to reconstruct the form of  $f(\bar{R})$. The metric in Jordan frame or modified gravity frame can be obtained by simply an inverse of the conformal transformation. The conformal factor $\Omega^2= f'=\Phi$  is the scalar in the modified frame. The scalar field in the Einstein frame $\phi$ is related to the $\Phi$ by $\phi=\frac{2}{\kappa}\ln\Omega=\frac{1}{\kappa}\ln\Phi$. The Ricci scalar in Einstein frame $\tilde{R}$ is related to the modified Ricci scalar$\bar{R}$ in the modified gravity through the expression given in \cite{2024EL....14729001K}. While $f'$ or $F=\frac{\partial f}{\partial\bar{R}}$ could be found directly from $\Phi$, $f(\bar{R})$ can be worked out from the expression of $V$, these expressions are obtained as function of space-time. In order to plot $f(\bar{R})$ we have gone through all the necessary transformations and plotted $f(\bar{R})$ and $f'(\bar{R})$ with respect to $t$ and $\bar{R}$ in figure (\ref{figf}). It is interesting to notice that, as time evolves from negative to positive, $f(\bar{R})$ and $f'(\bar{R})$ increase, reach a maxima at the bouncing epoch and bounce back following a similar trajectory. This agrees with our understanding that as universe moves away from the bouncing epoch both in positive and negative time direction, both $f(\bar{R})$ and $\bar{R}$ decrease. In other words both $f(\bar{R})$ and $\bar{R}$ are maximum at the bouncing epoch. The plot of $f'$ ensures that it is always positive, a requirement of the modified gravity theory. And clearly from the pattern of the plots, particularly $f'(\bar{R})$, one can infer that $f$ is not a linear function of $\bar{R}$

\begin{figure}[h!]
    \centering
    \caption{Plots of $f(\bar{R})$ and $f'(\bar{R})$ with respect to $\bar{R}$. The arrows on the blue trajectory show the flow of time from negative to positive. The red dot is the point of bounce occurring at $t=0$.}
    \includegraphics[height=3.5cm, width=0.48\linewidth]{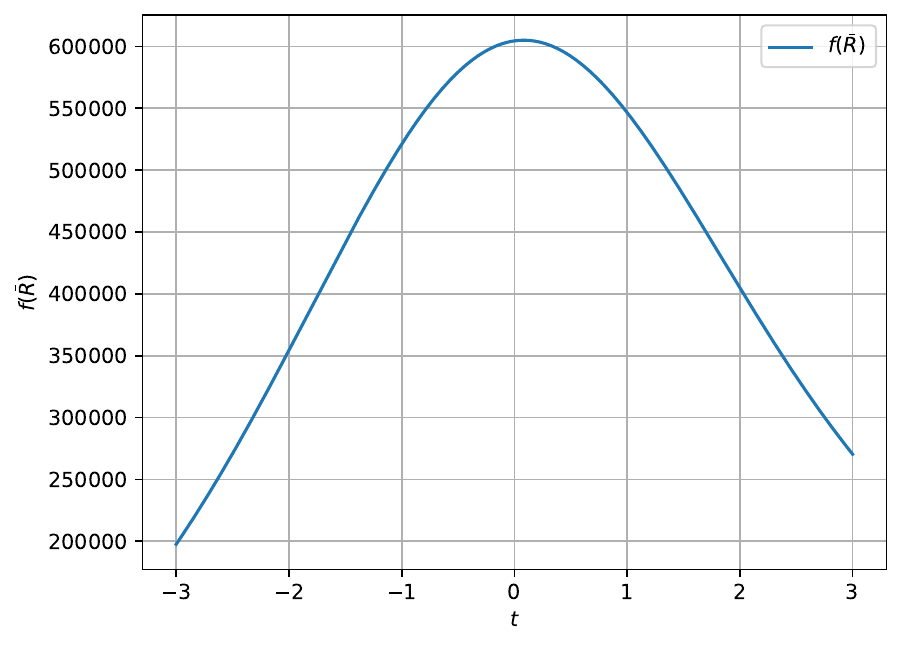}
    \includegraphics[height=3.5cm, width=0.48\linewidth]{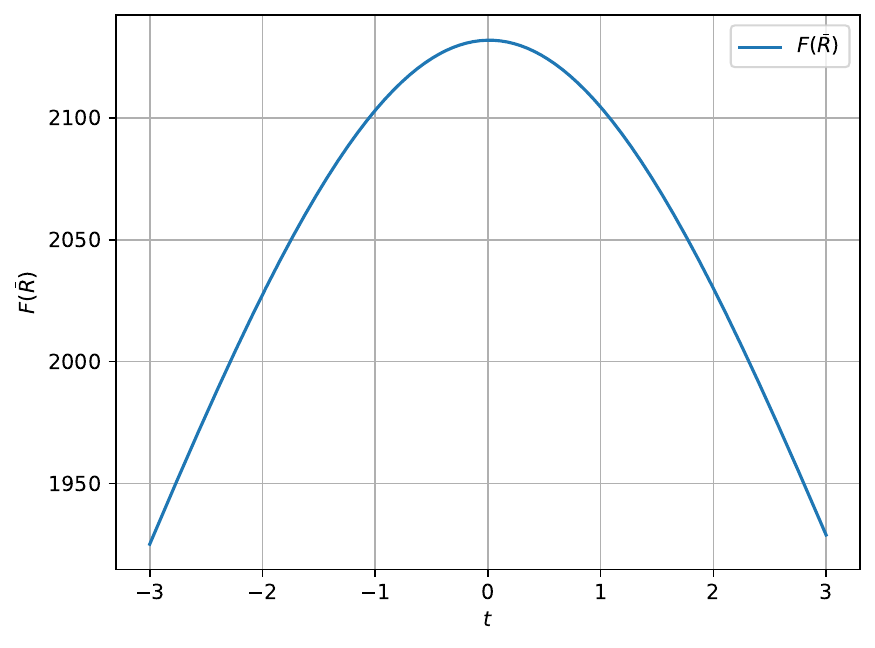}
     \includegraphics[height=3.5cm, width=0.48\linewidth]{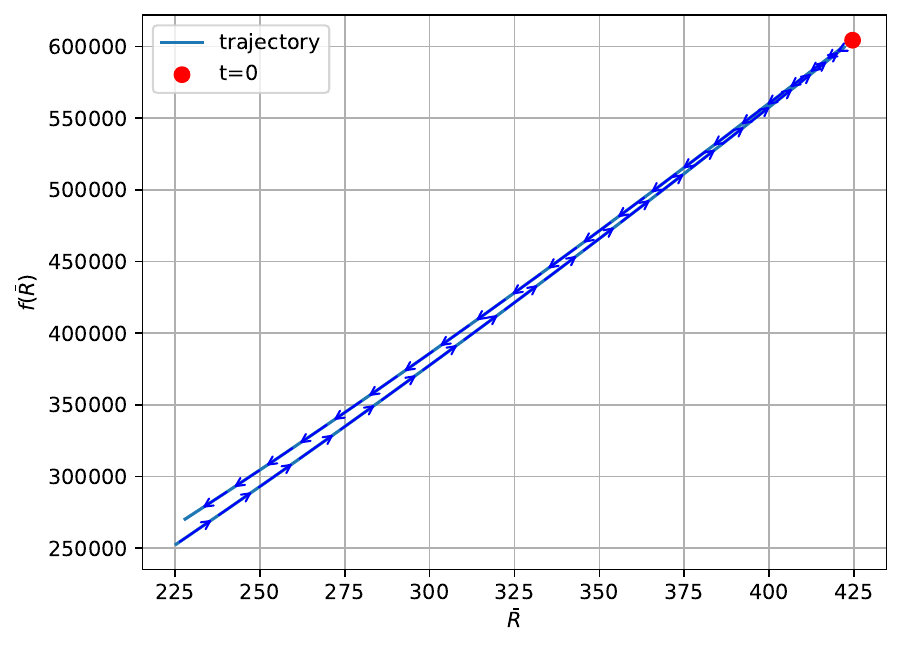}
     \includegraphics[height=3.5cm, width=0.48\linewidth]{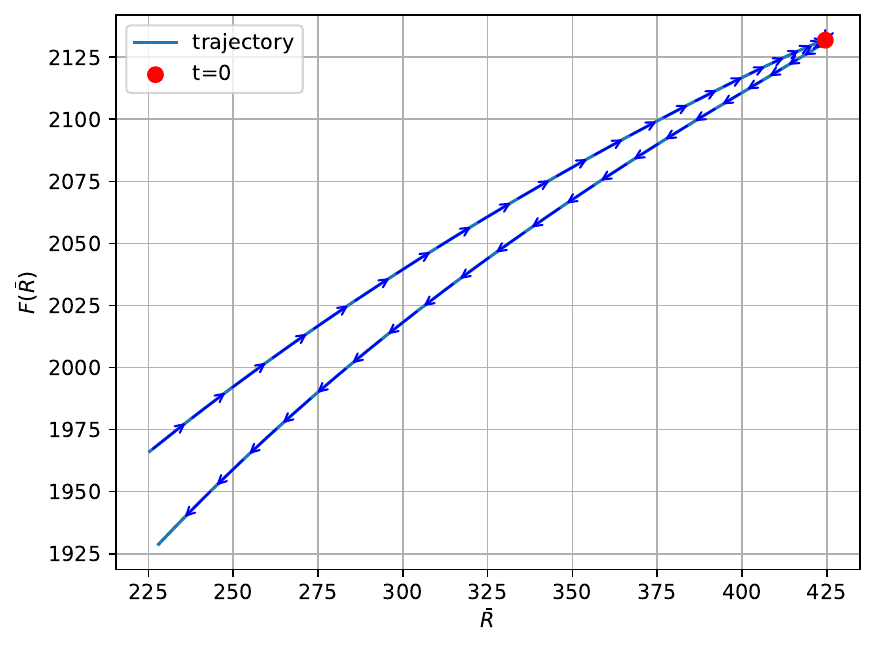}
    \label{figf}
\end{figure}
  


\item 

\textbf{Some comments on the perturbation of the fields and observational constraints:}

For the analysis to be complete, it is essential to investigate the behavior of the fields under perturbations and to impose appropriate constraints on the solutions based on observational data. As mentioned earlier, the torsion fields decay exponentially with time (\ref{a2p1}), giving the impression of a torsionless universe. This result also indicates that the bounce occurs only once in the cosmic evolution. So, we focus on the scalar field and analyse its behaviour under perturbation. The scalar field equation in the background of the inhomogeneous and anisotropic metric is 
\begin{equation}
\begin{aligned}
\square \phi
+ V_{,\phi}(\phi) - \frac{\kappa}{\sqrt{-g}} \partial_{\mu}\Big[\sqrt{-g}z_{ad}T^{\mu ad}
\Big] = 0 .
\end{aligned}
\label{bg-KG}
\end{equation}

Using the relations obtained earlier between different torsion field components and also the relations between derivatives of $\phi$, the above equation takes the form
\begin{equation}
\begin{aligned}
\ddot{\phi} + 3H\dot{\phi} 
- 3 g^{11}\,\partial_{1}^{2}\phi 
+ 3d\, g^{11}\,\partial_{1}\phi 
+ V_{,\phi}(\phi) \\
- \frac{6\kappa}{\sqrt{-g}} \Big[
\,\partial_{0}\!\left(\sqrt{-g}z_{13}T^{013}\right)
+ 3\,\partial_{1}\!\left(\sqrt{-g}\,z_{20}T^{012}\right)
\Big] = 0 .
\end{aligned}
\label{bg-KG}
\end{equation}
We consider scalar perturbations in Newton gauge, where the metric takes the form
\begin{equation}
ds^2 = -(1+2\Phi)\,dt^2
+ a^2(t)e^{-2d(x+y+z)}(1-2\Psi)\,\delta_{ij}dx^i dx^j .
\label{eq:newton-metric}
\end{equation}
and the scalar field is perturbed as
$\phi(t,\mathbf{x}) = \phi_0(t,\mathbf{x}) + \delta\phi(t,\mathbf{x}).$
In the Newtonian gauge, the perturbed  \cite{Unnikrishnan:2008qe,2020arXiv200704423T,PhysRevD.110.043533} scalar field ($\delta\phi$) equation is
\begin{equation}
\begin{aligned}
&\delta\ddot{\phi}
+ 3H\,\delta\dot{\phi}
-\frac{3}{a^{2} e^{-2d(x+y+z)}} 
\left(
\partial_{1}^{2}\delta\phi
- d\,\partial_{1}\delta\phi
\right)
+ V''(\phi_{0})\,\delta\phi= 
\dot{\phi}_{0}\,\dot{\Phi}+3\dot{\phi}_{0}\,\dot{\Psi}
- 2V'(\phi_{0})\,\Phi\\
&- \frac{3}{a^{2}e^{-2d(x+y+z)}} 
\left[ 2\Psi(\partial_{1}^{2}\phi_{0}- d\,\partial_{1}\phi_{0})-2\Phi(\partial_{1}^{2}\phi_{0}- d\,\partial_{1}\phi_{0})+ (\partial_{1}\Phi - \partial_{1}\Psi)\,\partial_{1}\phi_{0}\right]
\end{aligned}
\label{eq:pert-KG}
\end{equation}
The terms within the square bracket in the right hand side of equation (\ref{eq:pert-KG}) are the terms which arise due to the anisotrpy and inhomogeneity of the metric. These terms are $\sim {\cal{O}}(d^2)$ and higher. And since we divert very little from FRW universe and consider $d$ to be very small and hence these terms can be ignored. After a careful examination of all the terms in the equation, it can be concluded that the scalar perturbation follows a damped equation and exhibits a decaying behavior with time. Hence the solution of the scalar field remains stable under perturbation. 

Finally we relate our model with cosmological observations by 
constraining the model parameters with CMB distance observables. 
Our Hubble parameter $\tilde{H}$ has two parameters, $a_b$ and $A$(\ref{hnq}). We calculate the comoving distance $(r(z_*))$ at last scattering $z_*$  and the comoving sound horizon $(r_s(z_*))$ at $z_*$  using $\tilde{H}$. We further evaluate the CMB shift parameter $R$ and the acoustic scale $l_{A}$  and constrain the model parameters $a_b$ and $A$ by $\chi^2$ analysis using Planck data \cite{2019JCAP...02..028C}. Thus we find the allowed parameter space of $a_b$ and $A$ consistent with CMB constraints. 

\begin{figure*}[ht]
\centering
\includegraphics[height=5cm]{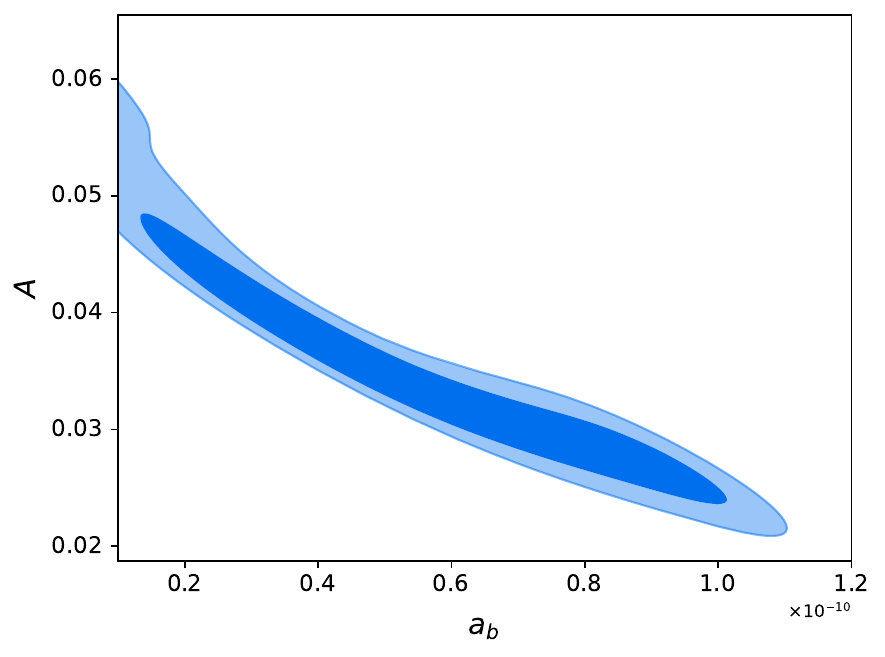}
\caption{Constraints on the parameters \(A\) and \(a_b\) from CMB distance priors (Planck 2018), showing the \(68\%\) and \(95\%\) confidence regions.}
\label{fig:wz_error}
\end{figure*}

\end{itemize}

\section{Conclusion}
The standard Big Bang Cosmological model suffer from an initial divergence problem where all the general laws of physics blow up. One of the many propositions to avoid this pathology, is  bouncing cosmology. To obtain a non-singular bounce, the minimum requirements are a vanishing $H$ and a positive $\dot{H}$ at the epoch of bounce and a decreasing and increasing $H$ before and after the bounce. To achieve a bounce, the null energy condition (NEC) gets violated in the frame work of GR with an isotropic homogeneous background. Two major hypotheses to achieve this are either by introducing new kind of matter or by modifying the frame work of Einstein gravity. Different kind of modification of gravity are available in the literature. The present work explores the scenario where Einstein gravity is modified through the spacetime connection to consider torsion, better known as Einstein Cartan theory. In modified gravity theories, the common trend is to carry the investigations in the modified frame with some functional form of curvature scalar. In these kind of studies, the field equations are separated into two parts- the Einstein gravity part and the modified gravity part, which is treated as the matter. The present work is not carried out in this fashion. We have not considered any functional form of the scalar $\bar{R}$, instead we reconstruct the form suitable for bounce once the minimum condition is met. Moreover, instead of separating the field equations, we move from the modified gravity frame to Einstein frame at the action level, and carry the investigations solely in Einstein frame, which is generally considered as the physical frame.  

In a generalized Einstein Cartan gravity or $f(\bar{R})$ gravity, torsion is incorporated in the spacetime manifold through the anti-symmetry of the affine connection. Thus the modified Riemann tensor contain terms responsible for torsion. $f(\bar{R})$ gravity action comprises of terms with modified Ricci scalar $\bar{R}$ constructed from the modified Riemann tensor. Following a similar methodology as in case of modified Einstein gravity $f(R)$ \cite{2010RvMP...82..451S,2024EL....14729001K}, modified Einstein Cartan gravity can be expressed as a non-minimal scalar tensor theory with the help of an auxiliary field and further, through conformal transformation of metric tensor, it can be mapped to an action of scalar-tensor theory minimally coupled to gravity in Einstein frame. The higher curvature and anti symmetric connection manifests in terms of  scalar ($\phi$) and  tensor ($Z_{ab}$) degrees of freedom which couples with each other through derivative coupling. A potential that depends on the form of $f(\bar{R})$ also appears for the scalar field .
Since the conformal transformation of the metric tensor recast the action in Einstein frame, the dynamics changes due to the changed background, differing from the studies in the modified frame.  

As $f(\bar{R})$ action is recast in the Einstein frame, apart from curvature, there are scalar field $\phi$ with a potential and three terms dependent on the torsion field $Z_{ab}$, which non-minimally couple with $\phi$ through derivative coupling. We perform our analysis in the context of an isotropic and homogeneous FRW universe, systematically introducing torsion-dependent terms in Cases 1, 2, and 3, defined by their respective coupling constants. In the FRW background, $\phi$ and $Z_{ab}$ fields depend on time only as they appear from the curvature. In all the three cases, we present the field equations which govern the dynamics. In all these cases, the field equations make it evident that a bouncing behavior is not possible, unless NEC is violated or the fields are phantom. The outcomes are similar to those found in the torsion-free case of an FRW universe within the framework of general relativity. Most of the torsion tensors vanishes due to the diktat of the off diagonal field equations. Due to symmetry of the spacetime, the curvature part in the off-diagonal field equations vanishes. These in turn force 3 of the 6 torsion components to vanish. In the first case, where only the first torsion dependent term is considered in the action (\ref{actfe}), only $T_{123}$ survives, suggesting the existence of only $Z_{12}$, $Z_{13}$ and $Z_{23}$. They can change the dynamics but are not enough to provide a bounce. In the second case, where the first two torsion-dependent terms in the action (\ref{actfe}) are included, the off-diagonal field equations yield a significant relation indicating that the non-vanishing components $Z_{12}$, $Z_{13}$ and $Z_{23}$ are proportional to  $\exp{(\kappa\phi)}$, linking them directly to the scalar field. In case 3, even after adding the last torsion-dependent term and considering the entire action (\ref{actfe}), the outcome is similar to case 2. Thus, interestingly in both cases 2 and 3, the antisymmetric tensors of rank 2 get identified with the scalar. This happens due to the constraints of the off-diagonal field equations. That is, the underlying spacetime symmetry governs whether certain components of the torsion tensor can exist, and does not favor the occurrence of a cosmological bounce.

This prompted us to relax the symmetry condition in space-time and go beyond the standard assumption of isotropy and homogeneity and understand the dynamics of the universe. A range of alternative cosmological models, such as Bianchi, Milne and others, explore the universe beyond standard cosmological assumptions. Various investigations have addressed bouncing cosmology in anisotropic universes, particularly within the framework of Bianchi models\cite{2022PhyS...97b5002S,2022ForPh..70a0065A,2018PhRvD..98b4009C}. Our purpose of relaxing the symmetry is to loosen up the stringent constraint condition imposed by the off-diagonal field equations and just explore such an alternative scenario. As we are exploring the phenomena in very early universe when gravity is very strong, we consider a toy metric which incorporates a deviation from the homogeneity and isotropic property of the metric through a spatially dependent scale factor. In the limit the deviation parameter $d$ tends to zero, the metric matches the FRW metric. We, in this work, consider $d$ to be very small to incorporate a marginal deviation from isotropy and homogeneity. In this anisotropic and inhomogeneous spacetime, both the scalar field and the torsion fields depend on space and time as they emerge from curvature. We study two cases (case 1 and 2) in this new spacetime based on the transformed action (\ref{actfe}) in Einstein frame. In this new cosmological setup, the scalar fields and torsion fields interact intricately with spacetime anisotropies and inhomogeneities, significantly affecting the effective dynamics.  The off diagonal field equations impose certain symmetry conditions on both scalar and torsion fields. These symmetry conditions constrain the torsion components but do not eliminate them. And even after all these, in the first case, where only the kinetic term of the torsion field is considered, it is not possible to get a bounce, like the earlier scenarios. 
In the second case however, the minimum requirement are met and a bouncing solution can be achieved without any phantom field. To find the exact solution, we consider a form of scale factor $\tilde{a}$ suitable to produce bounce and found the exact solutions for the scalar and the torsion fields that can be responsible for the bounce. The scalar field vary identically in all spatial directions.  We have presented a 3D  and a 2D plot of $\phi$, which show a maxima at the bouncing epoch. $V$ varies symmetrically around the maximum of $\phi$ at $t=0$. We also find the solutions for the torsion fields. Among them $Z_{01}$, $Z_{02}$ and $Z_{03}$ components are same and depends only on space, while the other three $Z_{12}$, $Z_{13}$ and $Z_{23}$ also have same solutions and depend on space and time both. Interestingly, the torsion field decays exponentially with time - a behavior consistent with its apparent absence from the observed universe. This also indicates that no more bounce can occur in future. Next, we reconstruct the  $f(\bar{R})$ by using inverse of the transformation and plotted both 
$f(\bar{R})$ and $f'(\bar{R})$ with respect to $\bar{R}$ and $t$. Both $f(\bar{R})$ and $f'(\bar{R})$ have maxima at the bouncing epoch. From these plots, we infer that the higher curvature is dominant around the bouncing epoch and the curvature decreases as the universe moves away from the bounce both in the positive and negative direction of time, which agrees with our general understanding. The framework presented here clearly favors a non-linear modification to standard gravity. Finally, we address the behaviour of the fields under  perturbation. As the torsion fields quickly decay, we construct the scalar perturbation in Newtonian gauge. An interesting thing to note is that the terms arising in the scalar perturbation equation due to anisotropy and inhomogeneity are very very small $\sim{\cal{O}}(d^2)$  and can be ignored. So the scalar perturbation follows a damped equation and decays with time, providing stability to the scalar solution under perturbation. We also constrained the model parameters $a_b$ and $A$ with Planck data, using $\chi^2$ analysis.

The findings indicate that, in the presence of torsion, higher-order curvature contributions within an inhomogeneous and anisotropic spacetime can generate a viable bounce and resolve singular behavior.
It opens new directions for exploring early-universe cosmology beyond standard paradigms and suggests that more realistic, less symmetric spacetime may naturally accommodate a bounce without exotic matter content.

\section{Acknowledgement}
SA acknowledges the CSIR fellowship provided by the Govt. of India under the CSIR-JRF scheme (file no. 09/0466(12904)/2021). SA and SS acknowledge the hospitality of IUCAA, Pune where part of the work is carried out.

\bibliographystyle{JHEP} 
\bibliography{main}


\section{Appendix 1}

 In this appendix we present the off diagonal field equations in the FRW background for the three different cases discussed in the main article.\\

\noindent
{\bf Case 1 :} Action (\ref{actfe1}) is considered in case 1 presented in section (3.1). The general field equations for action (\ref{actfe1}) are given by (\ref{fldeq1}). 
The off-diagonal field equations ($ij$) are
\begin{align}
    T_{031} T_{023} &= 0   &&\hspace{1cm} (\mu = 1,\ \nu = 2), \nonumber \\
    T_{012} T_{023} &= 0   &&\hspace{1cm} (\mu = 1,\ \nu = 3), \nonumber \\
    T_{012} T_{031} &= 0   &&\hspace{1cm} (\mu = 2,\ \nu = 3), \nonumber \\
    T_{023} T_{123} &= 0   &&\hspace{1cm} (\mu = 1,\ \nu = 0), \nonumber \\
    T_{031} T_{123} &= 0   &&\hspace{1cm} (\mu = 2,\ \nu = 0), \nonumber \\
    T_{012} T_{123} &= 0   &&\hspace{1cm} (\mu = 3,\ \nu = 0)
\end{align}
From these off-diagonal equations, it follows that in order to satisfy all constraints simultaneously, the only component of the torsion tensor that can remain non-zero is \( T_{123} \). All other components must vanish. \( T_{123} \) can be determined from the torsion field equation (\ref{c1t}).

\noindent
{\bf Case 2 :} Action (\ref{actfe2}) is considered in case 2 presented in section (3.2). The general field equations for action (\ref{actfe2}) are given by (\ref{fldeq2}). 
The off-diagonal components ($ij$) of the field equations are written as:
\begin{align}
(12)\ \rightarrow~ & \tilde{g}^{33}T_{023}\left[T_{031} - 2\kappa S_{031}\right] = 0, \\
(21)\ \rightarrow~ & \tilde{g}^{33}T_{031}\left[T_{023} - 2\kappa S_{023}\right] = 0, \\
(13)\ \rightarrow~ & \tilde{g}^{22}T_{023}\left[T_{012} - 2\kappa S_{012}\right] = 0, \\
(31)\ \rightarrow~ & \tilde{g}^{22}T_{012}\left[T_{023} - 2\kappa S_{023}\right] = 0, \\
(23)\ \rightarrow~ & \tilde{g}^{11}T_{031}\left[T_{012} - 2\kappa S_{012}\right] = 0, \\
(32)\ \rightarrow~ & \tilde{g}^{11}T_{012}\left[T_{031} - 2\kappa S_{031}\right] = 0,
\end{align}
\begin{align}
(01)\ \rightarrow~ & \tilde{g}^{22} \tilde{g}^{33}T_{023}\left[T_{123} - 2\kappa S_{123}\right] = 0, \\
(10)\ \rightarrow~ & \tilde{g}^{22} \tilde{g}^{33}T_{123}\left[T_{023} - 2\kappa S_{023}\right] = 0, \\
(02)\ \rightarrow~ & \tilde{g}^{11} \tilde{g}^{33}T_{031}\left[T_{123} - 2\kappa S_{123}\right] = 0, \\
(20)\ \rightarrow~ & \tilde{g}^{11} \tilde{g}^{33}T_{123}\left[T_{031} - 2\kappa S_{031}\right] = 0, \\
(03)\ \rightarrow~ & \tilde{g}^{11} \tilde{g}^{22}T_{012}\left[T_{123} - 2\kappa S_{123}\right] = 0, \\
(30)\ \rightarrow~ & \tilde{g}^{11} \tilde{g}^{22}T_{123}\left[T_{012} - 2\kappa S_{012}\right] = 0.
\end{align}
where $T_{ead}=\partial_{[e}Z_{ad]}$ and $S_{ead}=\partial_{[e}\phi Z_{ad]}$.
From the above set of off-diagonal equations, we observe that each equation contains a term in the square bracket of the form \( [T_{ead} - 2\kappa S_{ead}] \) multiplied with a torsion component. These equations suggest either the torsion components vanish or we get the following crucial relations 
\begin{equation}
T_{ead} = 2\kappa S_{ead}.
\end{equation}

\noindent
{\bf Case 3:}Action (\ref{actfe3}) is considered in case 3 presented in section (3.3). The general field equations for action (\ref{actfe3}) are given by (\ref{fldeq3}). 
The off-diagonal components ($ij$) of the above field equation yield the following constraints:
\begin{align}
(12)\ \rightarrow~ & T_{023} T_{031} - 2\kappa S_{031} T_{023} + \kappa^2 S_{031} S_{023} = 0, \\
(21)\ \rightarrow~ & T_{023} T_{031} - 2\kappa S_{023} T_{031} + \kappa^2 S_{031} S_{023} = 0, \\
(13)\ \rightarrow~ & T_{023} T_{012} - 2\kappa S_{012} T_{023} + \kappa^2 S_{012} S_{023} = 0, \\
(31)\ \rightarrow~ & T_{023} T_{012} - 2\kappa S_{023} T_{012} + \kappa^2 S_{012} S_{023} = 0, \\
(23)\ \rightarrow~ & T_{031} T_{012} - 2\kappa S_{012} T_{031} + \kappa^2 S_{012} S_{031} = 0, \\
(32)\ \rightarrow~ & T_{031} T_{012} - 2\kappa S_{031} T_{012} + \kappa^2 S_{012} S_{031} = 0,\\
(01)\ \rightarrow~ & T_{023} T_{123} - 2\kappa S_{123} T_{023} + \kappa^2 S_{123} S_{023} = 0, \\
(10)\ \rightarrow~ & T_{023} T_{123} - 2\kappa S_{023} T_{123} + \kappa^2 S_{123} S_{023} = 0, \\
(02)\ \rightarrow~ & T_{031} T_{123} - 2\kappa S_{123} T_{031} + \kappa^2 S_{123} S_{031} = 0, \\
(20)\ \rightarrow~ & T_{123} T_{031} - 2\kappa S_{123} T_{031} + \kappa^2 S_{031} S_{123} = 0, \\
(03)\ \rightarrow~ & T_{012} T_{123} - 2\kappa S_{123} T_{012} + \kappa^2 S_{123} S_{012} = 0, \\
(30)\ \rightarrow~ & T_{012} T_{123} - 2\kappa S_{012} T_{123} + \kappa^2 S_{123} S_{012} = 0.
\end{align}
From these equations, we get the following relations between $T_{ead}$ and $S_{ead}$
\begin{equation}
\frac{S_{031}}{T_{031}}  =  \frac{S_{023}}{T_{023}} = \frac{S_{012}}{T_{012}}  = \frac{ S_{123}}{T_{123}},
\end{equation}
After using these relations, we arrive at the following general relation:
\begin{equation}
T_{ead} T^{ead} - 2\kappa\, T_{ead} S^{ead} + \kappa^2 S_{ead} S^{ead} = 0.
\end{equation}

\section{Appendix 2}

In this appendix we present the off diagonal field equations in the inhomogeneous and anisotropic background given by metric (\ref{aim})for  different cases discussed in the main article.

\noindent
{\bf Case 1:} This case deals with action (\ref{actfe1}) in the inhomogeneous and anisotropic metric (\ref{aim}). The general field equations for action (\ref{actfe1}) are given by (\ref{fldeq1}). The diagonal equations are given in section 4.1. We list the off-diagonal field equations here. 
There are 6 off-diagonal ($ij$) equations out of which 3 have only space indices and 3 have space and time. As the spacetime is no longer isotropic and homogeneous, the curvature part of the off diagonal equations involving only space indices are non zero.
\begin{align}
(12)\rightarrow~&6\tilde{g}^{33}T_{023}T_{031} + \partial_1{\phi} \partial_2{\phi}= \frac{d^2}{\kappa^2}, \\ 
(31)\rightarrow~&6\tilde{g}^{22}T_{012}T_{023} + \partial_3{\phi} \partial_1{\phi}=\frac{d^2}{\kappa^2}\\
(23)\rightarrow~ &6\tilde{g}^{11}T_{031}T_{012} + \partial_2{\phi}\partial_3{\phi}=\frac{d^2}{\kappa^2}\\
(10)\rightarrow~ &6\tilde{g}^{22}\tilde{g}^{33}T_{123}T_{023} +\partial_0{\phi} \partial_1{\phi}=0,\\
(20)\rightarrow~  &6\tilde{g}^{11}\tilde{g}^{33}T_{123}T_{031} +\partial_0{\phi} \partial_2{\phi}=0 \\
(30)\rightarrow~ &6\tilde{g}^{11}\tilde{g}^{22}T_{123}T_{012} +\partial_0{\phi} \partial_3{\phi}=0.
\end{align}
From these equations it is straight forward to find that the scalar field satisfies the condition:
\begin{equation}   
\partial_1{\phi} = \partial_2{\phi} = \partial_3{\phi}
\end{equation}
and the torsion components follow  
\begin{equation}\label{a2tc1}
\frac{T_{012}}{T_{023}} =  \frac{T_{012}}{T_{031}}= \frac{T_{031}}{T_{023}}=1
\end{equation}
The relationship involving the $T_{123}$ with the other component  is expressed as:
\begin{equation}
\frac{T_{123}}{T_{023}}=\frac{T_{123}}{T_{012}}=\frac{T_{123}}{T_{013}}=\frac{\kappa^2\partial_0{\phi}\partial_1{\phi}}{\tilde{g}^{11}\left[\kappa^2(\partial_1{\phi})^2-d^2\right]}
\label{a2tcc}
\end{equation}

\noindent
{\bf Case 2:}
This case deals with action (\ref{actfe2}) in the inhomogeneous and anisotropic metric (\ref{aim}). The general field equations for action (\ref{actfe2}) are given by (\ref{fldeq2}). The diagonal equations are given in section 4.1. We list the off-diagonal field equations here. 
There are 12 off diagonal $(ij)$ field equations  - 6 of them involve only space indices and 6 involve time and space. 
\begin{align}
(12)\rightarrow~ & 6\tilde{g}^{33}T_{023}[T_{031} - 2\kappa S_{031}] + \partial_1{\phi} \partial_2{\phi}= \frac{d^2}{\kappa^2}, \\
(21)\rightarrow~ &6\tilde{g}^{33}T_{031}[T_{023} - 2\kappa S_{023}] + \partial_1{\phi} \partial_2{\phi}=\frac{d^2}{\kappa^2} , \\
(13)\rightarrow~ &6\tilde{g}^{22}T_{023}[T_{012} - 2 \kappa S_{012}] + \partial_3{\phi} \partial_1{\phi} =\frac{d^2}{\kappa^2} ,\\
(31)\rightarrow~ &6\tilde{g}^{22}T_{012}[T_{023} - 2 \kappa S_{023}] + \partial_3{\phi} \partial_1{\phi}=\frac{d^2}{\kappa^2} ,\\
(23)\rightarrow~ &6\tilde{g}^{11}T_{031}[T_{012} - 2\kappa S_{012}] + \partial_2{\phi} \partial_3{\phi}=\frac{d^2}{\kappa^2} ,\\
(32)\rightarrow~ &6\tilde{g}^{11}T_{012}[T_{012} - 2\kappa S_{012}] + \partial_2{\phi} \partial_3{\phi}=\frac{d^2}{\kappa^2} ,\\
(01)\rightarrow~ & 6\tilde{g}^{22}\tilde{g}^{33}T_{023}[T_{123} - 2\kappa S_{123}] +\partial_0{\phi} \partial_1{\phi}=0 ,\\
(10)\rightarrow~ & 6\tilde{g}^{22}\tilde{g}^{33}T_{123}[T_{023} - 2\kappa S_{023}] +\partial_0{\phi} \partial_1{\phi}=0,\\
(02)\rightarrow~ &6\tilde{g}^{11}\tilde{g}^{33}T_{031}[T_{123} - 2\kappa S_{123}] +\partial_0{\phi} \partial_2{\phi}=0,\\
(20)\rightarrow~ & 6\tilde{g}^{11}\tilde{g}^{33}T_{123}[T_{031} - 2\kappa S_{031}] +\partial_0{\phi} \partial_2{\phi}=0 ,\\
(03)\rightarrow~ &6\tilde{g}^{11}\tilde{g}^{22}T_{012}[T_{123} - 2\kappa S_{123}] +\partial_0{\phi} \partial_3{\phi}=0,\\
(30)\rightarrow~ &6\tilde{g}^{11}\tilde{g}^{22}T_{123}[T_{012} - 2 \kappa S_{012}] +\partial_0{\phi} \partial_3{\phi}=0.
\end{align}
From these off-diagonal equations, it is straightforward to find that the scalar field satisfies the following condition:
\begin{equation}   
\partial_1{\phi} = \partial_2{\phi} = \partial_3{\phi}
\end{equation}
which implies that the spatial dependence of $\phi$ is a $f(x+y+z)$ like in the previous case.
Also, the following $T_{ead}$ and $S_{ead}$ components obey the relations:
\begin{eqnarray}
\frac{T_{012}}{T_{023}} =  \frac{T_{012}}{T_{031}}= \frac{T_{031}}{T_{023}}=1
\label{a2tc1}\\
\frac{S_{012}}{S_{023}} =  \frac{S_{012}}{S_{031}}= \frac{S_{031}}{S_{023}}=1\label{a2sc1}
\end{eqnarray}
or in other words $T_{012}=T_{013}=T_{023}$ and $S_{012}=S_{013}=S_{023}$.
Finally, the relationship involving the $T_{123}$ with the other $T$ components and $S_{123}$ with the other $S$ components are expressed as:
\begin{eqnarray}
\frac{T_{123}}{T_{023}}=\frac{T_{123}}{T_{012}}=\frac{T_{123}}{T_{013}}=\frac{\kappa^2\partial_0{\phi}\partial_1{\phi}}{\tilde{g}^{11}\left[\kappa^2(\partial_1{\phi})^2-d^2\right]}
\label{a2tc}\\
\frac{S_{123}}{S_{023}}=\frac{S_{123}}{S_{012}}=\frac{S_{123}}{S_{013}}=\frac{\kappa^2\partial_0{\phi}\partial_1{\phi}}{\tilde{g}^{11}\left[\kappa^2(\partial_1{\phi})^2-d^2\right]}
\label{a2sc}
\end{eqnarray}

\end{document}